\documentclass[journal=jacsat,manuscript=article]{achemso}

\SectionNumbersOn

\usepackage[version=4]{mhchem} 
\usepackage{comment}
\usepackage{graphicx}
\usepackage{caption}
\usepackage{hyperref}
\usepackage{float}
\usepackage{booktabs}
\usepackage{multirow}
\usepackage{amsmath}
\usepackage{subcaption}
\usepackage[utf8]{inputenc}
\graphicspath{{Figures/}}
\author{Athul Tom\textsuperscript{a}}
\affiliation{Department of Chemical Sciences, Indian Institute of Science Education and Research Kolkata, Mohanpur, West Bengal, 741246, India}

\author{Pratyush Bhattacharjya\textsuperscript{a}}
\affiliation{Department of Chemical Sciences, Indian Institute of Science Education and Research Kolkata, Mohanpur, West Bengal, 741246, India}

\author{Sangita Sen}
\email{sangita.sen@iiserkol.ac.in}
\affiliation{Department of Chemical Sciences, Indian Institute of Science Education and Research Kolkata, Mohanpur, West Bengal, 741246, India}
\altaffiliation{Department of Chemical Sciences and Centre
for Advanced Functional Materials, Indian Institute of Science Education and Research (IISER) Kolkata, Mohanpur, West Bengal, 741246, India}

\title[DVR QD]
{Development of a Discrete Variable Representation of a Field-Adaptive Hermite Basis for Accurate Modeling of Magnetically Confined Systems}
\abbreviations{DVR,FD}

\AtBeginDocument{\footnotetext[1]{These authors have contributed equally to this work.}}

\begin{document}

\begin{abstract}
 Accurate magnetic-field evolution of spectra requires a common wavefunction representation that describes multiple states consistently across different field regimes. DVR methods combine the accuracy of basis-based evaluation of the kinetic-energy operator with the flexibility of a real-space representation of the potential, allowing arbitrary non-analytic potentials to be treated. In this paper, we develop a magnetic-field-dependent Hermite function-based (Hermite-B) Discrete Variable Representation (DVR) that naturally adapts to the magnetic confinement, providing an optimal representation of compressed wavefunctions over a wide range of field strengths while remaining centered in the regions most relevant to the scalar confinement potential. This ensures accurate treatment of the field-dependent kinetic energy and addresses two key challenges: maintaining comparable accuracy from the Zeeman to Landau regimes and providing flexibility for diverse confinement potentials. We also introduce a technique of spatially restricting the Hermite DVR grid points to a fixed-length domain, facilitating a compact basis representation for any smooth potential. The method is benchmarked for 2D quantum dots in a perpendicular magnetic field against analytically solvable models and compared with finite-grid, sine DVR, and field-free Hermite DVR calculations. The performance of the field-independent DVR methods is also investigated for magnetically confined systems for the first time. Excellent agreement with the exact analytical energy spectrum and wavefunctions is obtained using the DVR methods with the new Hermite-B basis requiring the fewest basis functions and hence a smaller Hamiltonian matrix, leading to computational efficiency. The applicability of our method is demonstrated for quantum dots under arbitrary confinement potentials and tilted magnetic fields, illustrating its versatility beyond analytically solvable systems. DVR-based methods, and in particular, the proposed field-dependent Hermite DVR could therefore provide a robust, accurate, and computationally efficient route towards large-scale calculations involving realistic semiconductor nanostructures as well as molecular systems in external magnetic fields.
\end{abstract}

\section{Introduction}
\label{sec:Introduction}
The accurate numerical solution of the Schrödinger equation remains one of the central challenges in quantum mechanics, particularly for systems subjected to external fields or complex confinement potentials where analytical solutions are generally unavailable\cite{SakuraiQM,ShankarQM,Tannor2007,Helgaker2000}. Such problems arise across atomic, molecular, and condensed-matter physics, including semiconductor quantum dots, ultracold trapped particles, excitons, and low-dimensional nanostructures.\cite{Jacak1998,Reimann2002,Hanson2007,Harrison2005} Reliable numerical methods capable of accurately describing these systems while maintaining computational efficiency are therefore of considerable importance.\cite{Leveque2007,Press2007,GolubVanLoan2013}

Among the numerous approaches developed for representing quantum mechanical wavefunctions, the Discrete Variable Representation (DVR)\cite{HarrisEngerholmGwinn1965,LightHamiltonLill1985,ColbertMiller1992,LightCarrington2000} has emerged as one of the most efficient in eigenvalue problems. DVR combines the advantages of basis-set expansions and grid-based methods by representing the wavefunction in an orthogonal basis while evaluating local operators on a discrete quadrature grid.\cite{LightHamiltonLill1985,ColbertMiller1992} As a result, the potential-energy operator becomes diagonal, significantly simplifying the construction of the Hamiltonian without sacrificing the spectral accuracy associated with basis-function methods.\cite{littlejohn2002multidimensional,LightCarrington2000} Owing to these properties, DVR techniques have found widespread applications in molecular spectroscopy, reaction dynamics, quantum scattering and quantum computing.\cite{BačićLight1986,MCTDH1990,brown2015using,asnaashari2026advantages,liu2022quantum,wang2009discrete,yurchenko2016duo,tennyson2004dvr3d,borrelli2016quantum,szidarovszky2010efficiency}

The efficiency of any DVR calculation, however, depends critically on the suitability of the underlying basis functions for the system.\cite{LightCarrington2000,ColbertMiller1992} Although conventional DVR formulations (sine, Hermite, etc.) are highly effective for systems whose characteristic length scales remain constant, they become increasingly inefficient when the spatial extent of the wavefunction varies significantly with external perturbations.\cite{Boyd2001,Canuto2006,shizgal2015spectral} In these situations, one often requires a substantially larger number of functions to achieve consistent accuracy, leading to increased computational cost and larger Hamiltonian matrices.\cite{LightHamiltonLill1985,LightCarrington2000}

Magnetically confined quantum systems provide a particularly important example of this challenge.\cite{Landau1930,Kohn1961,Maksym1990,Merkt1991} The application of an external magnetic field modifies both the effective kinetic energy and the confinement potential, leading to spatial localization of the wavefunctions.\cite{Landau1930,Kohn1961,Reimann2002} As the magnetic-field strength increases, the characteristic size of the wavefunction decreases owing to magnetic compression, while the corresponding energy spectrum evolves through the coupling between orbital motion and the magnetic field.\cite{Fock1928,Darwin1931,Maksym1990} The systems move from the field-free situation to the Zeeman, intermediate, and finally the Landau regime. For example, atoms which are spherical at $B=0$, become rugby-ball-shaped in the intermediate regime and finally needle-like in the Landau regime. Consequently, the optimal numerical representation of the wavefunction is itself expected to evolve with the magnetic field. A numerical representation that adapts naturally to the changing spatial characteristics of the system,\cite{LightCarrington2000} is expected to facilitate computations in such situations.

Semiconductor GaAs quantum dots constitute an ideal model system for investigating these effects.\cite{Bastard1988,Davies1998,Harrison2005,Jacak1998}. Quantum dots exhibit discrete and highly tunable energy spectra due to quantum confinement\cite{Jacak1998}. The application of external magnetic fields introduces additional control over these energy levels through mechanisms such as Zeeman splitting and orbital coupling, leading to the emergence of rich physical phenomena such as the Fock-Darwin spectrum\cite{Reimann2002}. This tunability makes quantum dots ideal platforms for studying fundamental quantum mechanics and applications in quantum computing, optoelectronics, and nanoscale sensing\cite{Kouwenhoven2001,loss1998quantum,bimberg1999quantum}. Electrons confined in semiconductor quantum dots are often modeled using a two-dimensional harmonic oscillator potential due to the approximately parabolic nature of the confinement\cite{ikhdair2012spectra,que1992excitons,niculescu2001energy,quang2021strongly,voskoboynikov2001spin}. Experimental measurements have directly observed harmonic oscillator-like wavefunctions in quantum dots, validating this approximation\cite{teichmann2013harmonic}. Their approximately parabolic confinement gives rise to the well-known Fock--Darwin spectrum for a single electron subjected to a perpendicular magnetic field, providing one of the few quantum-confined systems for which exact analytical solutions are available.\cite{Fock1928,Darwin1931,Landau1930,Madhav1994,Kouwenhoven2001,Reimann2002} Because both the energy eigenvalues and wavefunctions are known analytically, the Fock--Darwin model offers a rigorous benchmark for assessing the accuracy and convergence of numerical methods.\cite{nikhily2025,mukherjee2026modelling,Merkt1991,Reimann2002} Furthermore, by introducing non-parabolic confinement potentials or arbitrarily oriented magnetic fields, the same framework can be extended to physically realistic situations where analytical solutions no longer exist and robust numerical approaches become essential.\cite{bednarek2003modeling,peeters1990magneto,dippel1994charged}

The DVR method, which is popular in the field of quantum dynamics of atomic and molecular systems has received very little attention for quantum dots. The sine DVR method has been explored for field-free 2D GaAs quantum dots \cite{prudente2005study,aytekin2013nonlinear} and mean-field basis-based DVR has been employed for many-electron quantum dots\cite{duron2017computation} in a few isolated studies. In the present work, we develop a magnetic-field-dependent Hermite (Hermite-B) DVR in which the scaling of the Hermite basis is determined by the effective magnetic confinement. The methodology is first validated through detailed comparisons with analytical Fock-Darwin solutions as well as finite-grid, sine DVR, and field-independent Hermite DVR calculations. Its applicability is subsequently demonstrated for quantum dots with arbitrary confinement potentials and for systems subjected to arbitrarily oriented magnetic fields, illustrating that the formulation provides a general and computationally efficient framework for treating quantum-confined systems in complex electrostatic and magnetic environments. To the best of our knowledge, this work presents the first exploration of DVR methods for systems in magnetic fields.


\section{Theory}
\label{sec:Theory}
This section presents the theoretical framework underlying the proposed field-dependent Hermite discrete variable representation (DVR) for two-dimensional quantum systems subjected to external magnetic fields. We first discuss the Hamiltonian for the 2D system subjected to a perpendicular magnetic field and the analytic Fock-Darwin solution corresponding to an isotropic harmonic scalar confinement potential. The Fock-Darwin solution provides us with the effective confinement frequency which is used in the field-dependent Hermite basis which is developed in this paper. This is followed by the details of how the DVR for the selected sets and the corresponding grids are generated. Subsequently, we discuss how this method can be applied to anisotropic harmonic, arbitrary (non-harmonic) scalar potentials as well as magnetic fields of arbitrary orientation (not perpendicular).

\subsection{2D Model Hamiltonian with Perpendicular Magnetic Field}
\label{sec:modelH}
The motion of a single electron confined to two-dimensions can be described within the effective-mass ($\mu$) approximation,\cite{Bastard1988,Davies1998,Harrison2005} in which the influence of the potential on the kinetic energy is incorporated through an effective electron mass. Within this approximation, the non-relativistic Hamiltonian of an electron in a quantum dot subjected to an external magnetic field $\mathbf{B}$ can be introduced through the minimal-coupling prescription,

\begin{equation} \hat{H} = \frac{1}{2\mu} \left( \hat{\mathbf p} - q\hat{\mathbf A} \right)^2 + \hat{V}(\hat{x},\hat{y}), \label{minimalc} \end{equation}

where $\mu$ and $q$ denote the effective mass and charge of the particle, respectively, $\hat{\mathbf A}$ is the vector potential associated with the external magnetic field, and $\hat{V}(\hat{x},\hat{y})$ represents the confinement potential. In the present formulation, $V(x,y)$ is kept completely general, allowing the Hamiltonian to describe arbitrary two-dimensional confinement potentials. This general form constitutes the starting point for the numerical discrete variable representation (DVR) treatment.

The DVR basis is developed from the system where a magnetic field is applied perpendicular to the plane of confinement,\cite{Landau1930,Fock1928,Darwin1931}

\begin{equation} \mathbf{B}=B\hat{\mathbf z}, \label{eq:B_z} \end{equation}

for which the vector potential in the symmetric gauge is adopted,

\begin{equation} \hat{\mathbf A} = \frac{B}{2} (-\hat y,\hat x,0), \label{gauge1} \end{equation}

since it preserves the rotational symmetry of the perpendicular-field problem.\cite{LandauLifshitzQM,SakuraiQM} Substituting Eq.~(\ref{gauge1}) into the Hamiltonian of Eq.~(\ref{minimalc}) gives

\begin{equation} \begin{aligned} \hat H = & \frac{1}{2\mu} \left( \hat p_x^2+\hat p_y^2 \right) + \frac{qB}{2\mu} \left( \hat y\hat p_x - \hat x\hat p_y \right) + \frac{q^2B^2}{8\mu} \left( \hat x^2+\hat y^2 \right) + \hat{V}(\hat{x},\hat{y}). \label{eq:HOeq} \end{aligned} \end{equation}

Equation~(\ref{eq:HOeq}) retains the general confinement potential $V(x,y)$ and therefore does not, in general, possess an analytical solution. The generality of this Hamiltonian is particularly important for the present work, where the DVR formulation, post-benchmarking, is applied to confinement potentials that need not admit an analytically solvable Schrodinger equation.

\subsection{Diagonalization-Based Discrete Variable Representation}

The numerical methodology developed in this work is based on the diagonalization-based discrete variable representation (DVR)\cite{LightHamiltonLill1985,ColbertMiller1992,littlejohn2002multidimensional}. In this formulation, the kinetic energy operator is evaluated in the finite basis representation (FBR), whereas the potential-energy operator is evaluated in the discrete variable representation (DVR). The resulting approach exploits the advantages of both representations: the kinetic-energy operator retains its compact analytical form in the FBR, while the local potential-energy operator becomes diagonal in the DVR, considerably simplifying the structure of the Hamiltonian.

The diagonalization of the Hamiltonian is done in FBR. The transformation between the FBR and DVR is obtained by diagonalizing the coordinate operator matrix represented in the finite basis, $\mathbf{X}$ which is diagonalized as
\begin{equation} \mathbf{X} = \mathbf{U Q U^\dagger}, \label{eq:diag_X} \end{equation}

where $\mathbf{U}$ is the unitary transformation matrix whose columns contain the eigenvectors of $\mathbf{X}$, and $\mathbf{Q}$ is the corresponding diagonal coordinate operator in the DVR. For basis functions generated from orthogonal polynomials, the coordinate operator satisfies a three-term recurrence relation, resulting in a real symmetric tridiagonal matrix in the FBR.\cite{LightHamiltonLill1985,littlejohn2002multidimensional} This tridiagonal structure is fundamental to the DVR construction, since its eigenvalues and eigenvectors define the Gaussian quadrature associated with the chosen basis. Specifically, the eigenvalues correspond to the quadrature nodes (DVR grid points), while the eigenvectors determine the transformation between the FBR and DVR and the corresponding quadrature weights.\cite{LightCarrington2000,golub1969calculation,golub2007calculation}. Thus, the eigenvalues of the coordinate matrix define the DVR grid points.
Consequently, matrix elements of any local multiplicative operator are evaluated directly at the quadrature nodes, yielding
\begin{equation} \mathbf{Q_{\alpha\beta}} =x_\alpha \delta_{\alpha\beta}, \label{eq:Q_DVR} \end{equation}

where $x_\alpha$ denotes the coordinate associated with the $\alpha$th DVR grid point.
The associated Gaussian quadrature approximates coordinate-space integrals according to
\begin{equation} \int_a^b f(x) \omega(x) dx\approx \sum_{\alpha=1}^N \omega_\alpha f(x_\alpha), \label{eq:gauss_quad} \end{equation}

where $x_\alpha$ and $\omega_\alpha$ denote the quadrature nodes and weights, respectively. The quadrature is exact for polynomials up to the degree supported by the underlying orthogonal basis and ensures the discrete orthogonality and completeness relations satisfied by the DVR basis functions.\cite{LightCarrington2000} Thus, any local potential operator becomes diagonal in the DVR basis,

\begin{equation} \mathbf{V^{\mathrm{DVR}}_{\alpha\beta}} = V(x_\alpha)\delta_{\alpha\beta}, \label{eq:V_DVR} \end{equation}

which is one of the defining features of the DVR formalism allowing the Hamiltonian matrix to be constructed as

\begin{equation} \mathbf{ H^{\mathrm{FBR}} = T^{\mathrm{FBR}} + U V^{\mathrm{DVR}} U^\dagger}, \label{eq:dvr_hamiltonian} \end{equation}

where $\mathbf{T^{\mathrm{FBR}}}$ denotes the kinetic-energy operator in the finite basis representation.

This formulation is independent of the particular choice of basis functions and therefore represents one of the principal strengths of the DVR methods.\cite{ColbertMiller1992,LightCarrington2000} Once the coordinate and kinetic-energy operators are specified in a chosen basis, the remaining DVR construction proceeds similarly. Consequently, both the sine and Hermite DVR schemes considered in this work employ the same underlying numerical framework, differing only in the basis used to represent the coordinate and kinetic-energy operators. This flexibility makes the diagonalization-based DVR particularly attractive for developing adaptive basis representations tailored to specific physical systems.

\subsection{Sine DVR}
\label{sec:Sine_DVR}

As a first numerical representation, we employ the popular sine DVR, which uses the eigenfunctions of the one-dimensional particle-in-a-box Hamiltonian as the finite basis representation. Owing to their orthogonality, simple analytical structure, and natural satisfaction of hard-wall boundary conditions, sine basis functions have become one of the most widely adopted choices in DVR calculations.\cite{ColbertMiller1992,prudente2001discrete,richman2026utilizing,yurchenko2016duo,piserchia2015discrete,asnaashari2026advantages}

For a particle of mass $\mu$ in a box of length $L$, the one-dimensional basis functions are

\begin{equation} \varphi_n(x)= \sqrt{\frac{2}{L}} \sin\left[ \frac{n\pi}{L} \left( x+\frac{L}{2} \right) \right], \quad n=1,2,3,\ldots \label{eq:sine_basis} \end{equation}

Within this basis, the kinetic (T)-, momentum (P)-, and position (X)-operator matrices assume simple analytical forms,

\begin{equation} \mathbf{T}_{mn} = \frac{\hbar^2}{2\mu} \left( \frac{n\pi}{L} \right)^2 \delta_{mn}, \label{eq:T_sine} \end{equation}

\begin{equation} \mathbf{P}_{mn} = \begin{cases} -\dfrac{i\hbar}{L} \dfrac{4mn}{m^2-n^2}, & m+n\ \mathrm{odd}, \\[8pt] 0, & \mathrm{otherwise}, \end{cases} \label{eq:P_sine} \end{equation}

and

\begin{equation} \mathbf{X}_{mn} = \begin{cases} -\dfrac{2L}{\pi^2} \left[ \dfrac{1}{(m-n)^2} - \dfrac{1}{(m+n)^2} \right], & m+n\ \mathrm{odd}, \\[10pt] 0, & m+n\ \mathrm{even}. \end{cases} \label{eq:X_sine} \end{equation}

Unlike the Hermite basis introduced in the following subsection, the coordinate (position) operator is not tridiagonal in the sine basis. Consequently, the DVR grid is constructed by diagonalizing a transformed coordinate operator,

\begin{equation} f(x) = \cos \left[ \frac{\pi}{L} \left( x+\frac{L}{2} \right) \right], \label{eq:sine_transform} \end{equation}

The matrix representation of the transformed coordinate $\mathbf{F}_{jk}$ is tridiagonal and is given by,

\begin{equation} \mathbf{F}_{jk} = \frac12 \left( \delta_{j,k+1} + \delta_{j,k-1} \right). \label{eq:F_ij} \end{equation}

Diagonalization of $\mathbf{F}$ gives the eigenvalues $f_\alpha$ and eigenvectors $\mathbf{U}_\alpha$. $\mathbf{U}_\alpha$ is the unitary transformation matrix and the grid points $x_\alpha$ can be obtained from $f_\alpha$ through,

\begin{equation} x_\alpha = -\frac{L}{2} + \frac{L}{\pi} \cos^{-1}(f_\alpha). \label{eq:gp_sine} \end{equation}

The two-dimensional basis is obtained by forming the direct product of the corresponding one-dimensional functions,

\begin{equation} \Phi_{mn}(x,y) = \varphi_m(x) \otimes \varphi_n(y), \label{eq:2D_basis_sine} \end{equation}

leading to the finite basis Hamiltonian

\begin{align} \mathbf{H_{m'n',mn}^{\mathrm{FBR}}} &= \mathbf{T}_{m'm}\otimes \mathbf{I}_{n'n} + \mathbf{I}_{m'm}\otimes \mathbf{T}_{n'n} \nonumber\\ &\quad + \frac{qB}{2\mu} \left( \mathbf{ X}_{m'm}\otimes \mathbf{P}_{n'n} - \mathbf{P}_{m'm}\otimes \mathbf{X}_{n'n} \right) + \mathbf{V^{\mathrm{FBR}}_{m'n',mn}}, \label{eq:H_DVR_sine} \end{align}

where $\mathbf{I}$ denotes the identity matrix. The potential-energy contribution is obtained through the standard DVR transformation. 

\begin{equation} \mathbf{V^{\mathrm{FBR}}} = (\mathbf{U}_x\otimes \mathbf{U}_y) \mathbf{V^{\mathrm{DVR}}} (\mathbf{U}_x\otimes \mathbf{U}_y)^\dagger, \label{eq:V_FBR} \end{equation}
and 

\begin{equation} \mathbf{V}^{DVR}_{\alpha'\beta',\alpha\beta}=V(x_\alpha,y_\beta)\delta_{\alpha\alpha'}\delta_{\beta\beta'} \label{eq:diag_DVR_2D} \end{equation}

where, $x_\alpha$ and $y_\beta$ are grid points, and $\mathbf{U}_x$ and $\mathbf{U}_y$ the matrices of eigenvectors of $\mathbf{F}$, each represented along the x and y directions. The sine DVR provides a well-established comparator for assessing the performance of the proposed field-dependent Hermite DVR developed in the following subsection.
\subsection{Field-Dependent Hermite DVR}
\label{sec:Hermite_DVR}

 To allow the basis to adapt to the progressive localization of the electronic wavefunction as the magnetic confinement increases thereby avoiding the possibility of higher errors at higher field strengths, we introduce a magnetic-field-dependent Hermite DVR in which the basis functions and the associated DVR grid evolve naturally with the effective confinement.

The basis functions of FBR are constructed from the eigenfunctions of the one-dimensional quantum harmonic oscillator. Owing to their orthogonality, completeness, and direct connection with harmonic confinement, Hermite functions constitute a natural basis for quantum systems possessing approximately parabolic potentials.\cite{Szego1939,boyd1984asymptotic} Unlike the conventional Hermite DVR, however, the present formulation introduces a magnetic-field-dependent scaling parameter $\alpha$ in the one-dimensional basis functions,

\begin{equation} \varphi_n(x)= \frac{1}{\sqrt{2^nn!}} \left( \frac{\alpha}{\pi} \right)^{1/4} H_n(\alpha x) e^{-\alpha^2x^2/2}, \qquad n=0,1,2,\ldots, \label{eq:her_exp} \end{equation}

where, $H_n$ are Hermite polynomials. To obtain the field-dependent parameter $\alpha$ for the Hermite-B basis, we specialize the confinement potential of  Eq.~(\ref{eq:HOeq}) to the two-dimensional isotropic harmonic form,

\begin{equation} V_{\mathrm{har}}(x,y) = \frac{1}{2}\mu\omega_0^2 \left( x^2+y^2 \right), \label{harmonic_potential} \end{equation}

where $\omega_0$ is the characteristic confinement frequency, leading to the well-known Fock--Darwin Hamiltonian,\cite{Fock1928,Darwin1931,Madhav1994,Kouwenhoven2001,Reimann2002} for which the eigenenergies are given by the Fock--Darwin spectrum,

\begin{equation} E_{n,M} = \left( 2n+|M|+1 \right) \hbar\Omega - \frac{1}{2} \hbar\omega_c M, \label{Energy_expression} \end{equation}

where $n=0,1,2,\ldots$ and $M=0,\pm1,\pm2,\ldots$ are the radial and magnetic quantum numbers, respectively. The so-called cyclotron frequency $\omega_c $ is defined as

\begin{equation} \omega_c = \frac{qB}{\mu}, \label{cyclotron_frequency} \end{equation}

with the sign of $q$ retained explicitly, while the effective confinement frequency becomes

\begin{equation} \Omega = \sqrt{ \omega_0^2 + \frac{\omega_c^2}{4} }. \label{effective_frequency} \end{equation}
The scaling parameter $\alpha$ depends explicitly on the applied magnetic field through $\Omega$, 
\begin{equation}
\alpha
=
\sqrt{\frac{\mu\Omega}{\hbar}},
\label{eq:alpha}
\end{equation}
allowing the basis to evolve continuously as the magnetic confinement changes. However, these are not the analytic Fock-Darwin eigenfunctions which involve Laguerre polynomials. 
The analytical Fock--Darwin eigenfunctions can be written in terms of the radial quantum number $n$ and magnetic quantum number $M$ in the polar form as

\begin{equation} \Psi_{n,M}(r,\phi) = \frac{1}{\ell_B} \sqrt{ \frac{n!} {\pi (n+|M|)!} } \left( \frac{r}{\ell_B} \right)^{|M|} e^{-r^2/(2\ell_B^2)} L_n^{|M|} \left( \frac{r^2}{\ell_B^2} \right) e^{iM\phi}, \label{FD_wavefunction_polar} \end{equation}

where $L_n^{|M|}$ denotes the associated Laguerre polynomial and $\ell_B$ is the magnetic-field-dependent effective oscillator length,

\begin{equation} \ell_B = \sqrt{\frac{\hbar}{\mu\Omega}}. \label{FD_length} \end{equation}

and in Cartesian coordinates as

\begin{equation} \Psi_{n,M}(x,y) = \frac{1}{\ell_B} \sqrt{ \frac{n!} {\pi (n+|M|)!} } \left( \frac{\sqrt{x^2+y^2}}{\ell_B} \right)^{|M|} \exp\left[ -\frac{x^2+y^2}{2\ell_B^2} \right] L_n^{|M|} \left( \frac{x^2+y^2}{\ell_B^2} \right) \left( \frac{x+iy}{\sqrt{x^2+y^2}} \right)^M. \label{FD_wavefunction_cartesian} \end{equation}
where

\begin{equation} r=\sqrt{x^2+y^2}, \qquad e^{iM\phi} = \left( \frac{x+iy}{r} \right)^M, \label{cartesian_polar_relation} \end{equation}

with the appropriate interpretation for negative $M$.

For $M<0$, the angular factor is equivalently written as

\begin{equation} \left( \frac{x+iy}{r} \right)^M = \left( \frac{x-iy}{r} \right)^{|M|}. \label{eq:conjugate_M} \end{equation}

The above expression provides the analytical reference wavefunctions directly on the Cartesian $(x,y)$ grid used in the DVR calculations, while retaining the natural $(n,M)$ quantum-number labeling of the Fock--Darwin states. The corresponding probability density is independent of the phase and is given by

\begin{equation} |\Psi_{n,M}(x,y)|^2 = \frac{1}{\ell_B^2} \frac{n!} {\pi (n+|M|)!} \left( \frac{x^2+y^2}{\ell_B^2} \right)^{|M|} \exp\left[ -\frac{x^2+y^2}{\ell_B^2} \right] \left[ L_n^{|M|} \left( \frac{x^2+y^2}{\ell_B^2} \right) \right]^2. \label{FD_probability} \end{equation}

In the Fock-Darwin eigenstates the magnetic field modifies the effective oscillator frequency through $\Omega$, but the Cartesian harmonic-oscillator functions provide a convenient basis for representing the analytical solution. Appropriate linear combinations of the degenerate Cartesian states $\Phi_{n_x,n_y}(x,y)$ will diagonalize $\hat L_z$. The analytical Fock--Darwin energies and corresponding wavefunctions serve as the primary reference for assessing the convergence, accuracy, and efficiency of the numerical DVR formulations developed in the following sections. However, it is important to note that only the effective confinement frequency $\Omega$ is defined analogous to that in the Fock--Darwin model while the field-dependent basis developed here uses Hermite polynomials. Physically, this adaptive scaling reflects the magnetic compression of the electronic wavefunction and is closely related to adaptive spectral basis techniques developed for field-free quantum mechanical calculations as well.\cite{Boyd2001,Canuto2006}

The two-dimensional basis functions are obtained from the direct product of the corresponding one-dimensional Hermite functions, similar to the Sine DVR functions, 

\begin{equation} \Phi_{mn}(x,y) = \varphi_m(x) \otimes \varphi_n(y). \label{eq:basis_hermite} \end{equation}

A particularly attractive feature of the Hermite basis is that the coordinate operator possesses a tridiagonal matrix representation,

\begin{equation} \mathbf{X}_{mn} = \frac{1}{\sqrt{2}\alpha} \left( \sqrt{n}\, \delta_{m,n-1} + \sqrt{n+1}\, \delta_{m,n+1} \right), \label{eq:X_hermite} \end{equation}

which enables the DVR grid to be obtained directly through diagonalization of the coordinate matrix without introducing an auxiliary coordinate transformation, as required in the sine DVR. This is called the tridiagonal quadrature rule.\cite{dickinson1968calculation}

Within the same basis, the kinetic-energy operator is given by

\begin{equation} \mathbf{T}_{mn} = \frac{\hbar\Omega}{4} \left[ (2n+1)\delta_{mn} - \sqrt{n(n-1)} \delta_{m,n-2} - \sqrt{(n+1)(n+2)} \delta_{m,n+2} \right], \label{eq:T_hermite} \end{equation}

while the orbital angular momentum operator assumes the matrix representation

\begin{equation} \mathbf{L}_{m'm,n'n}^{(z)} = -i\hbar\left(\sqrt{n'(n+1)} \delta_{m',n'-1} \delta_{m,n+1} - \sqrt{n(n'+1)} \delta_{m',n'+1} \delta_{m,n-1}\right). \label{eq:L_z_matrix} \end{equation}

The Hamiltonian in the finite basis representation therefore becomes

\begin{equation} \mathbf{H_{m'n',mn}^{\mathrm{FBR}}} = \mathbf{T}_{m'm}\otimes \mathbf{I}_{n'n} + \mathbf{I}_{m'm}\otimes \mathbf{T}_{n'n} + \frac{\omega_1}{2}\mathbf{L}^{(z)}_{m'm,n'n} + \mathbf{V^{\mathrm{FBR}}_{m'n',mn}}, \label{eq:H_FBR_hermite} \end{equation}

where, $\mathbf{I}$ is the identity matrix.
The potential-energy matrix is obtained using the same DVR transformation introduced in Section~\ref{sec:Sine_DVR}. Consequently, the overall numerical framework remains identical to that of the sine DVR, with the only difference lying in the choice of basis functions used to represent the operators.

\section{Numerical Implementation}
\label{sec:NumericalImplementation}

All numerical calculations were performed using in-house Python programs
developed with the NumPy and SciPy scientific-computing libraries.
\cite{Harris2020NumPy,Virtanen2020SciPy} The numerical implementation follows
the theoretical formulation described in Section~\ref{sec:Theory}. The
different representations were implemented within a common computational
framework, with the basis-specific coordinate, momentum, and kinetic-energy
matrices constructed according to the corresponding formulations described
above.

The kinetic-energy, coordinate, and momentum operators were constructed analytically in the finite basis representation. The magnetic field-dependent terms were then assembled from the corresponding Cartesian operators. For the DVR calculations, the coordinate matrices were diagonalized to obtain the DVR grid points and the associated transformation matrices. The coordinate-dependent confinement potential was evaluated directly at the DVR grid points and represented as a diagonal matrix. The resulting potential matrix was transformed to the finite basis representation before being combined with the remaining Hamiltonian terms. This procedure was applied for both the sine and Hermite DVR formulations.

The two-dimensional Hamiltonian was constructed from the corresponding one-dimensional matrices using Kronecker products as shown in Eqs. \ref{eq:H_DVR_sine} and \ref{eq:H_FBR_hermite}. For a basis containing $N$ functions along each Cartesian coordinate, the resulting two-dimensional Hamiltonian has dimension $N^2\times N^2$. The Hamiltonian matrices were diagonalized using the eigensolvers provided
by SciPy.\cite{Virtanen2020SciPy} The dense DVR Hamiltonians were treated as
Hermitian matrices and diagonalized using the corresponding dense Hermitian
eigensolver. The finite-difference Hamiltonian was represented as a sparse
matrix and the required low-lying eigenpairs were obtained using the sparse
Hermitian eigensolver.

At each magnetic-field strength, the lowest twenty eigenpairs were retained for the field-dependent calculations. The eigenvalues were arranged in ascending order and the corresponding normalized eigenvectors were retained for subsequent analysis. To maintain a consistent correspondence between eigenstates at successive magnetic-field values, an overlap-based state-tracking procedure was used. For two consecutive field values, the absolute overlaps between the corresponding sets of normalized eigenvectors were evaluated. An energy dependent weighting was included together with the overlap criterion, and
the resulting assignment problem was solved using the Hungarian algorithm.
\cite{Kuhn1955,Munkres1957}The procedure was applied successively over the magnetic-field range to assign corresponding eigenstates between neighboring field values. The state-tracking procedure was used only for establishing the correspondence between eigenstates and did not modify the Hamiltonian diagonalization.

\section{Results and Discussion}
\label{sec:Results}

We compare four representations: uniform Cartesian grid, Sine DVR, field-independent Hermite-0 DVR, and field-dependent Hermite-B DVR. Performance is assessed through basis requirements, spectral and wavefunction accuracy, and computational cost. Sine uses the finite-box basis of Section~\ref{sec:Sine_DVR}; both Hermite variants use the oscillator basis of Section~\ref{sec:Hermite_DVR}. Hermite-0 retains its zero-field scaling, whereas Hermite-B updates its scale using the effective confinement frequency at each field. For Woods--Saxon calculations, the spatial extent is selected from the potential parameters and basis/grid dimensions (Section~\ref{sec:WS_parameters}).

The discussion proceeds from the analytically solvable isotropic GaAs benchmark to anisotropic InAs confinement (with an analytical normal-mode reference), finite-depth Woods--Saxon confinement (without an analytical magnetic-field solution), and tilted fields. This sequence tests convergence, spectra, grid and wavefunction representations, and the response to continuously varying effective perpendicular confinement.

\subsection{Isotropic Harmonic Confinement: GaAs Quantum Dot}

The GaAs calculations cover $B=0$--$5$~T with effective mass $\mu=0.067m_e$ and confinement frequency $\omega_0=32.26$~cm$^{-1}$. The Grid reference uses a uniform $512\times512$ Cartesian mesh with 5~a.u. spacing; Sine uses a box length of 2560~a.u.; and Hermite-0 and Hermite-B use, respectively, fixed zero-field and field-dependent oscillator scaling. The lowest twenty eigenstates were calculated at each field and compared with the Fock--Darwin solution, using basis sizes established by the convergence studies.

\subsubsection{Basis-Set Convergence}

Convergence of the lowest ten eigenstates was defined by an energy change below $10^{-2}$~cm$^{-1}$ between successive basis sizes.

\begin{figure}[H]
\centering
\includegraphics[width=\linewidth]{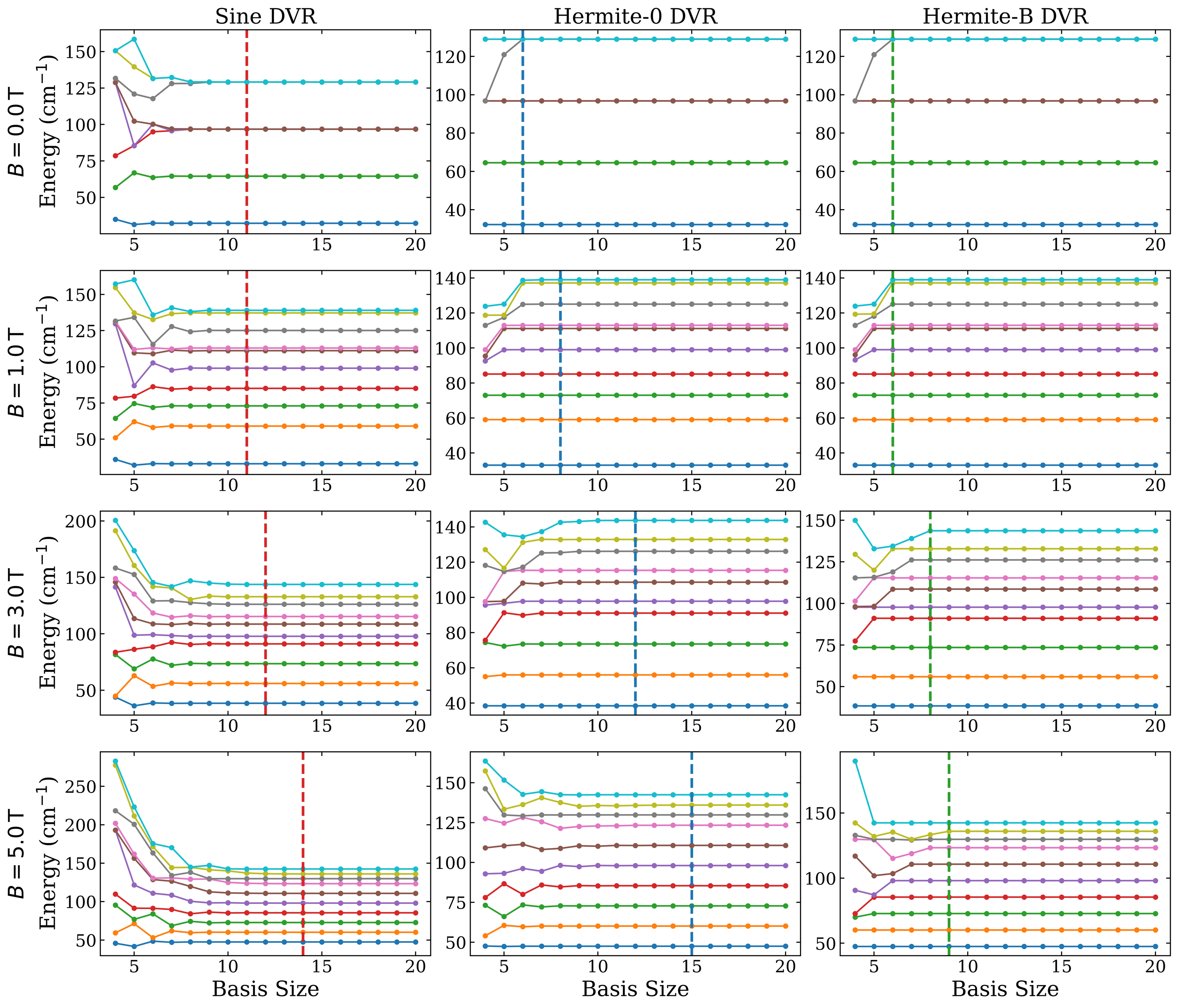}
\caption{Basis-set convergence of the lowest ten eigenstates obtained using the sine DVR (left column), Hermite-0 DVR (middle column), and Hermite-B DVR (right column) for representative magnetic fields of $B=0$, 1, 3, and 5 T. The dashed vertical lines indicate the minimum basis size satisfying the convergence criterion of $10^{-2}$ cm$^{-1}$ between succesive iterations.}
\label{grand_fig}
\end{figure}

At $B=0$, 1, 3, and 5~T, $N_{\rm conv}=11,11,12,14$ for Sine; $6,8,12,15$ for Hermite-0; and $6,6,8,9$ for Hermite-B (Fig.~\ref{grand_fig}). Hermite-0 is efficient at low fields but loses efficiency as magnetic compression increases; Hermite-B requires the fewest functions throughout.

\begin{figure}[H]
\centering
\includegraphics[width=0.6\linewidth]{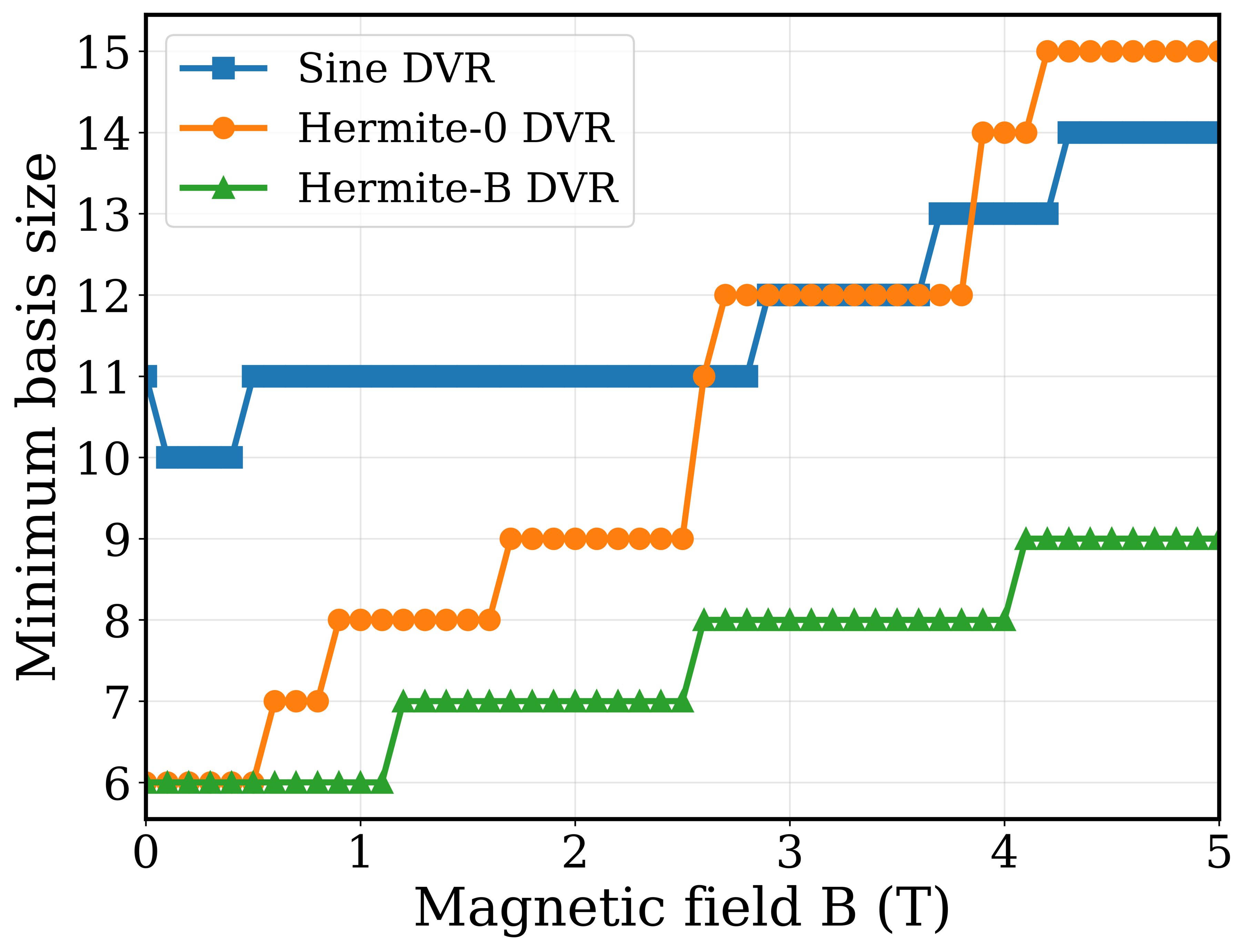}
\caption{Comparison of the basis sizes required to achieve an energy convergence threshold of $10^{-2}$ cm$^{-1}$ for the sine DVR, Hermite-0 DVR, and Hermite-B DVR.}
\label{fig:basis_compare_ho}
\end{figure}

Figure~\ref{fig:basis_compare_ho} summarizes the trend of the converged basis size with field strength. The computational advantage of Hermite-B would become valuable in higher-dimensional and many-electron calculations, where matrix dimensions grow rapidly with basis size.

\subsubsection{Energy Spectrum and Benchmarking Against the Fock--Darwin Model}

Production DVR spectra were compared directly with the Fock--Darwin solution using $N=12$ functions per coordinate, a basis size sufficient for convergence of all the methods considered.

\begin{figure}[H]
\centering
\includegraphics[width=0.8\linewidth]{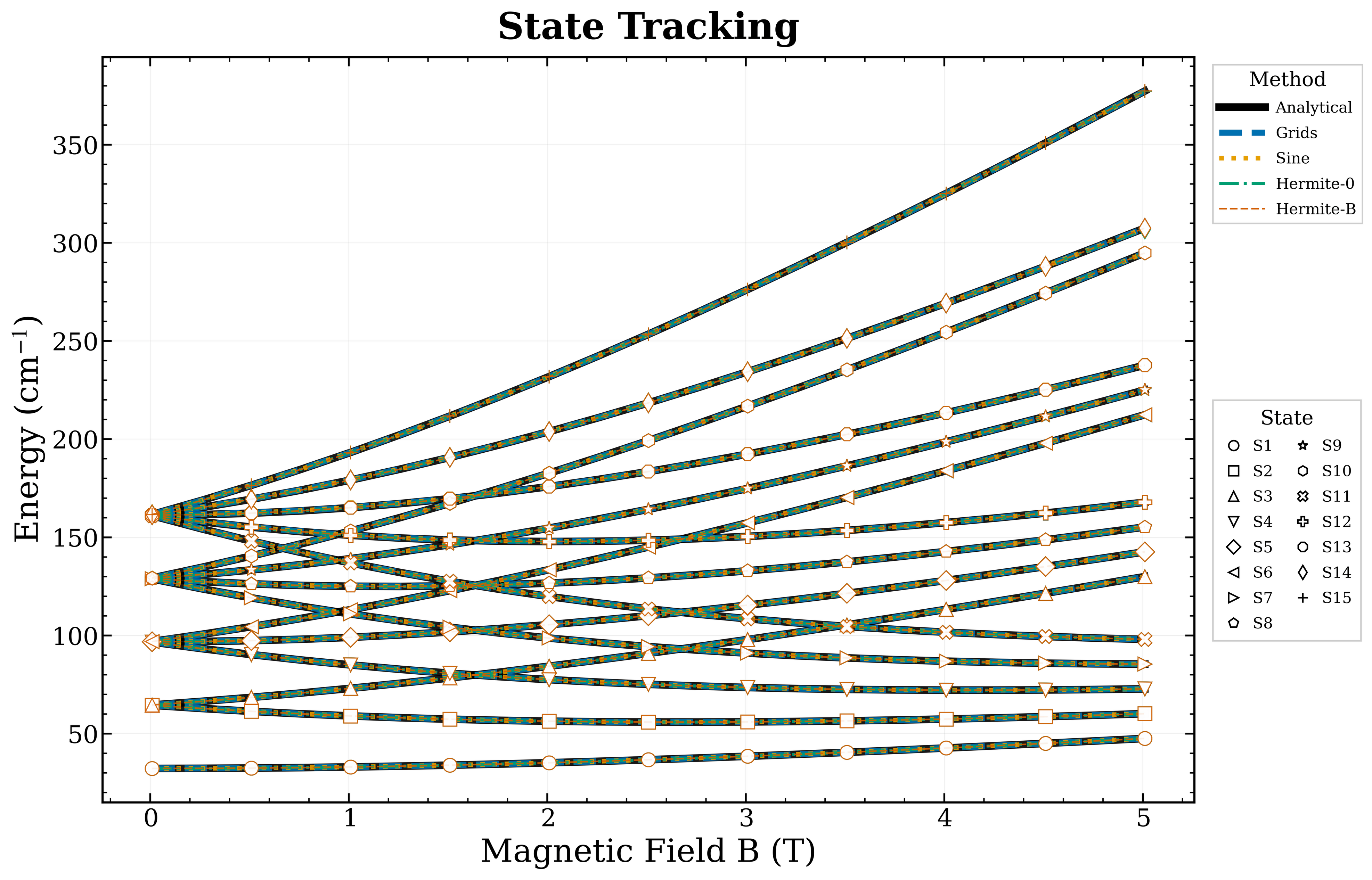}
\caption{Evolution of the fifteen lowest energy states of the two-dimensional GaAs quantum dot as a function of magnetic field. The analytical Fock--Darwin energies are compared with the corresponding results obtained using the finite-grid, sine DVR, Hermite-0 DVR, and Hermite-B DVR approaches. The numerical results closely follow the analytical spectrum over the entire magnetic-field range.}
\label{fig:ho_full_spectrum}
\end{figure}

\begin{figure}[H]
\centering
\includegraphics[width=0.8\linewidth]{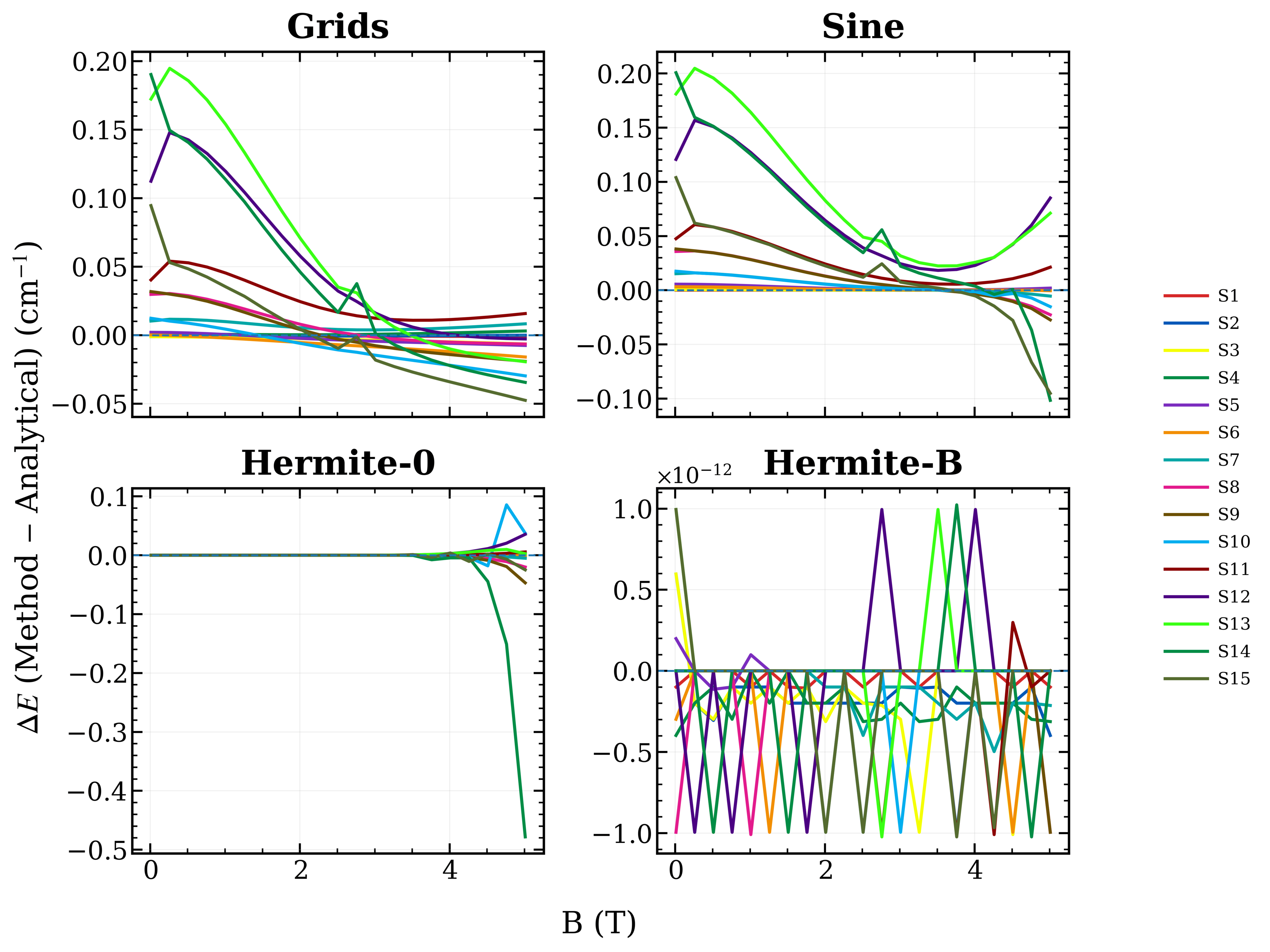}
\caption{Deviations of the first fifteen energy states (in cm$^{-1}$) obtained using the finite-grid, sine DVR, Hermite-0 DVR, and Hermite-B DVR approaches from the analytical Fock--Darwin energies as a function of magnetic field.}
\label{fig:error_methods_ho}
\end{figure}

Figure~\ref{fig:ho_full_spectrum} shows the fifteen lowest levels. Orbital coupling progressively lifts the zero-field degeneracies among states with equal $2n+|M|$, producing the Fock--Darwin fan and changing state ordering. All four numerical spectra nearly coincide with the analytical curves; Fig.~\ref{fig:error_methods_ho} resolves their differences.

Grid and Sine have the largest errors (up to $0.2$~cm$^{-1}$ for excited states at low fields). These generally decrease with $B$, but at 5~T state-specific drifts reach about $-0.05$~cm$^{-1}$ (Grid) and $-0.1$~cm$^{-1}$ (Sine); Sine is more accurate for lower states. Both use the same 2560~a.u. box, selected to contain extended excited states with a fine mesh~\cite{nikhily2025}. Their difference therefore reflects the representation, although box-truncation and discretization errors cannot be separated without a box-size study. Hermite-0 errors are negligible to about 3.5~T and generally below $0.05$~cm$^{-1}$ thereafter, with one state reaching $-0.5$~cm$^{-1}$ near 5~T. Hermite-B remains at approximately $10^{-12}$~cm$^{-1}$ without discernible field or state dependence.

\subsubsection{DVR Grid Structure}
\label{sec:DVR_grid}
\begin{figure}[H]
\centering
\includegraphics[width=\linewidth]{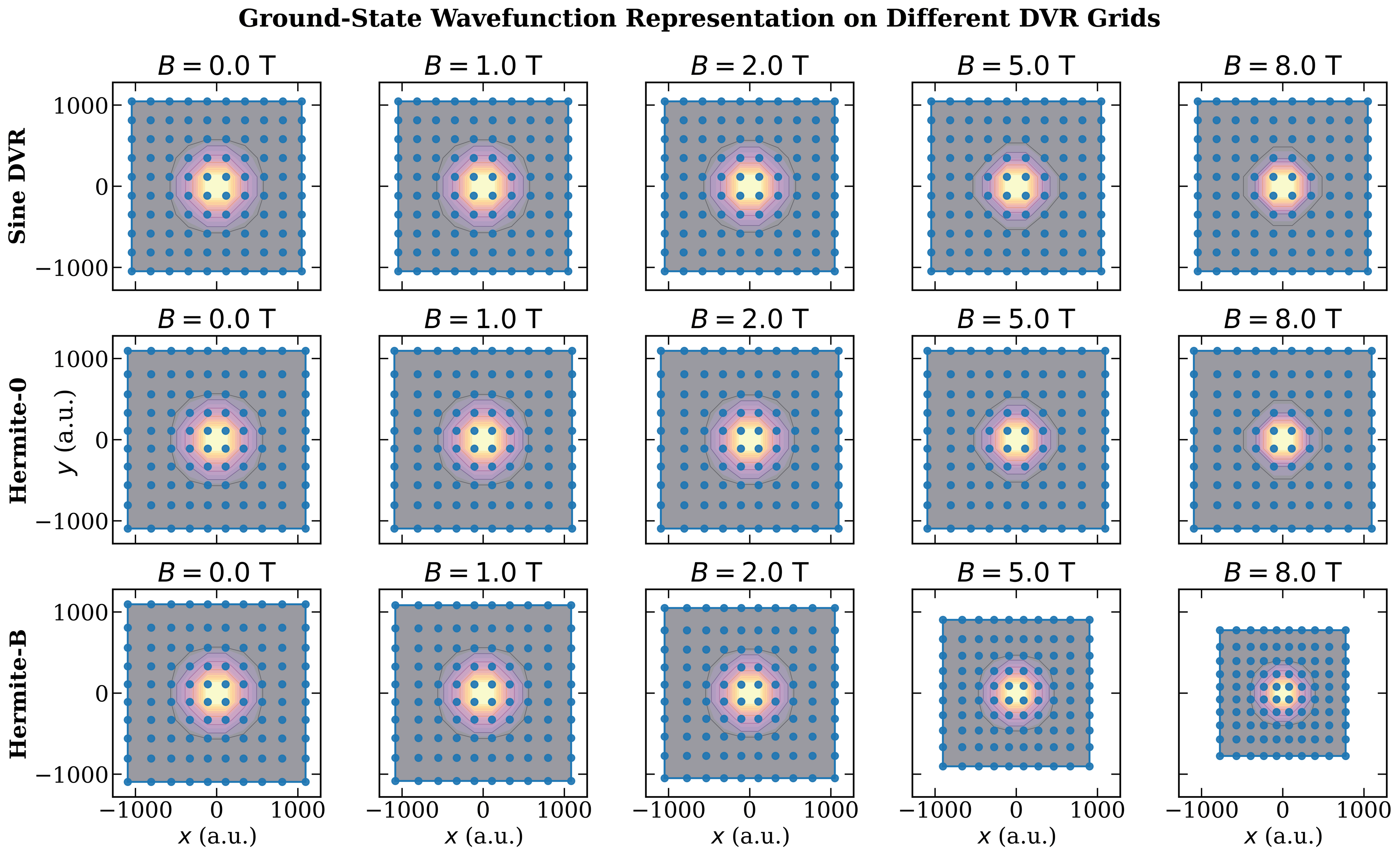}
\caption{Comparison of the ground-state wavefunction representation and spatial distribution of the DVR grid points for the sine DVR, Hermite-0, and Hermite-B at representative magnetic-field strengths. The calculations use $N_x=N_y=10$ basis functions.}
\label{fig:grid_compare}
\end{figure}

The distinct convergence patterns are also reflected in the spatial distributions of the DVR grids (Fig.~\ref{fig:grid_compare}).

The uniform Sine grid does not adapt its sampling density to magnetic localization. Hermite-0 is non-uniform and well matched at $B=0$, but its fixed oscillator length leaves the compressed wavefunction occupying a progressively smaller grid region. Hermite-B coincides with Hermite-0 at zero field and contracts with increasing confinement and follows the probability density better. For the displayed $N_x=N_y=10$, this redistributes rather than removes points: more sample the high-probability region and fewer the negligible-amplitude tails.

\subsubsection{Wavefunction Analysis}
\label{sec:wavefn}
Wavefunctions complement energy benchmarks by testing spatial structure, probability densities, transition amplitudes, and magnetic localization.

The root-mean-square extent of each centered and normalized state $\psi_i(x,y)$ was evaluated as
\begin{equation} \Sigma_R = \sqrt{\sigma_x^2+\sigma_y^2}. \label{eq:sigma_r} \end{equation}

Where
\begin{equation} \sigma_x = \sqrt{ \langle x^2\rangle-\langle x\rangle^2 }, \qquad \sigma_y = \sqrt{ \langle y^2\rangle-\langle y\rangle^2 }. \end{equation}

As, $\langle x\rangle=\langle y\rangle=0$, so
\begin{equation} \Sigma_R=\sqrt{\langle r^2\rangle}, \qquad r^2=x^2+y^2. \end{equation}

From the Fock--Darwin eigenfunctions, with oscillator length $l_B$ (Eq.~\ref{FD_length}), the analytical radial moment is
\begin{equation} \langle r^2\rangle_{n,M} = l_B^2 \left(2n+|M|+1\right), \end{equation}
and the corresponding reference spread is
\begin{equation} \Sigma_R^{\mathrm{FD}} = l_B \sqrt{2n+|M|+1}. \label{eq:fd_spread} \end{equation}

\begin{figure}[H]
    \centering

    \begin{subfigure}{0.6\textwidth}
        \centering
        \includegraphics[width=\textwidth]
        {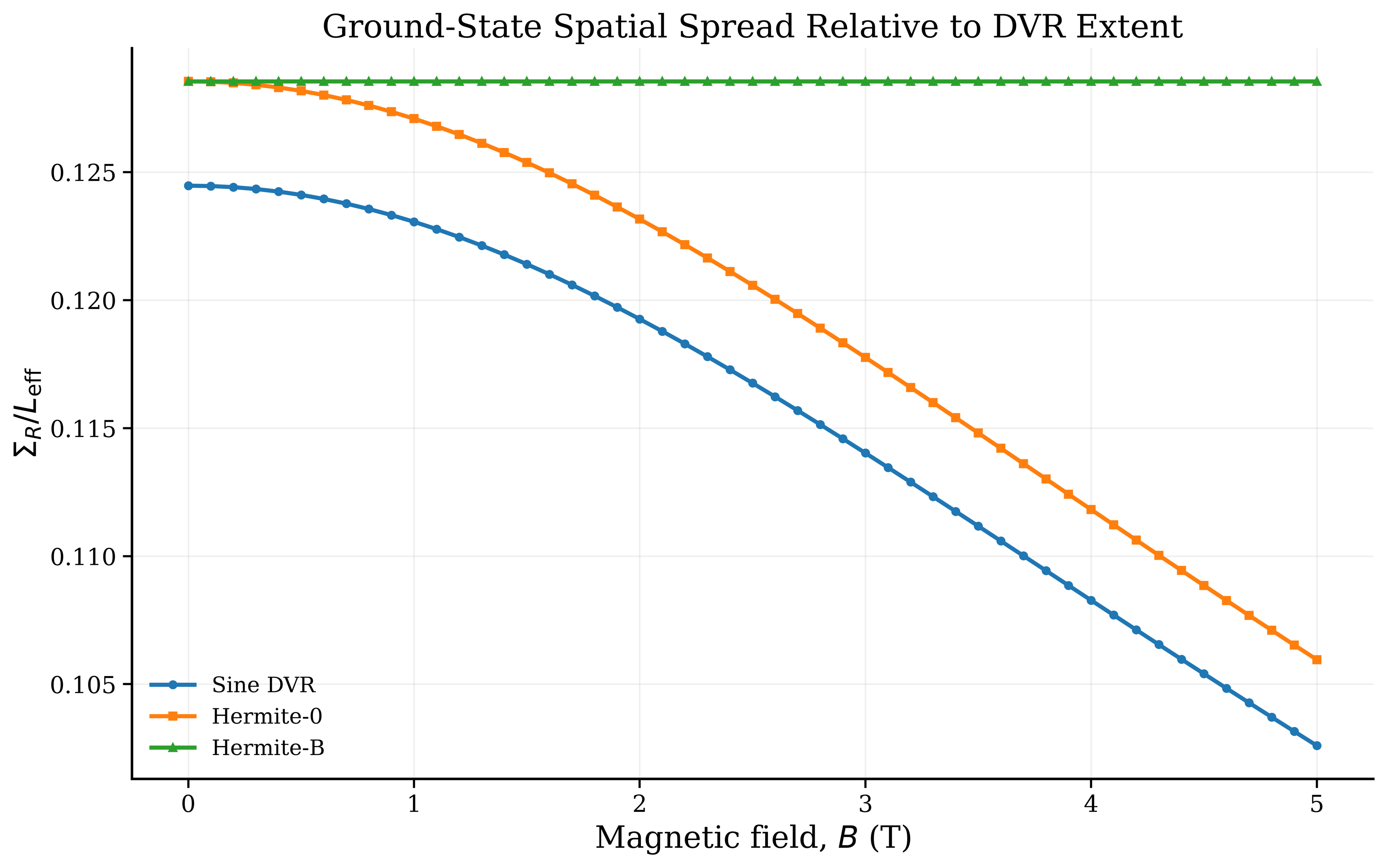}
    \end{subfigure}
    \vspace{0.5cm}

    \begin{subfigure}{\textwidth}
        \centering
        \includegraphics[width=\textwidth]
        {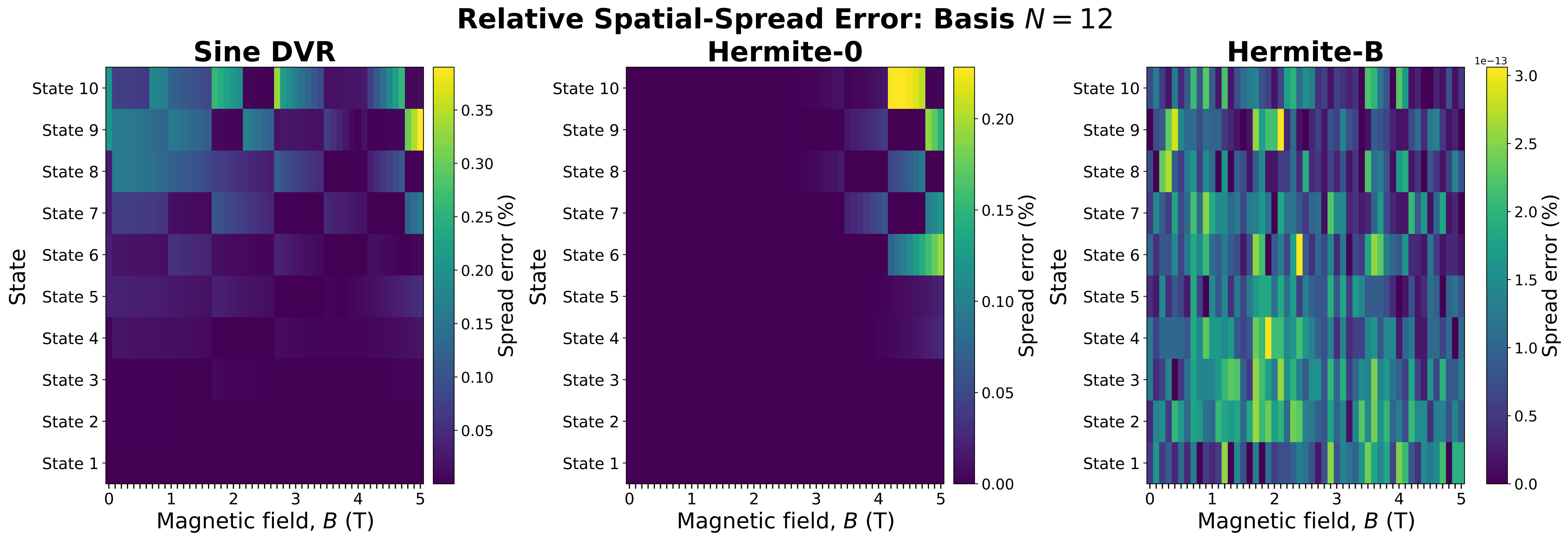}
    \end{subfigure}
 
    \caption{(a) Ground-state spatial spread, $\Sigma_R/L_{\mathrm{eff}}$, versus magnetic-field strength. The Sine and Hermite-0 ratios decrease with field, while Hermite-B remains approximately constant. (b) Relative spatial-spread error, $\epsilon_{\Sigma_R}$, for the first ten states at $N=12$ for the Sine, Hermite-0, and Hermite-B DVRs.}
    
    \label{fig:gs_sigma_l}
\end{figure}

Here, $L_{\mathrm{eff}}$ is the DVR-grid extent computed from the grid points furthest apart. Figure~\ref{fig:gs_sigma_l}(a) compares ground-state $\Sigma_R/L_{\mathrm{eff}}$ at the common $N=12$ basis, isolating representation quality from minimum-basis effects. The ratios are similar at $B=0$; with increasing field they decrease for Sine and Hermite-0 but remain nearly constant for Hermite-B. Smaller bases further favor Hermite-B.

In Figure~\ref{fig:gs_sigma_l}(b), the same $N=12$ basis is used to compare the first ten states through the relative error 

\begin{equation} \epsilon_{\Sigma_R} = 100 \frac{ \left| \Sigma_R^{\mathrm{DVR}} - \Sigma_R^{\mathrm{FD}} \right| }{ \Sigma_R^{\mathrm{FD}} }. \label{eq:sigma_error} \end{equation}

 as heatmaps (field and state index on the horizontal and vertical axes) for Sine, Hermite-0, and Hermite-B. Sine has the broadest, state-dependent deviations, largest for excited states. Hermite-0 is accurate for most low-lying states at low/intermediate fields, but develops localized high-field errors, especially for excited states. Hermite-B remains near $10^{-13}\%$ for all ten states and fields, with only numerical noise and no systematic field- or state-dependent growth. Note that, the panels are independently normalized with their colour bars giving maxima of approximately $0.4\%$ (Sine), $0.2\%$ (Hermite-0), and $10^{-13}\%$ (Hermite-B).


\subsection{Anisotropic Harmonic Confinement: InAs Quantum Dot}
\label{sec:InAs_results}
\label{sec:InAs_theory}

The anisotropic InAs dot tests Hermite-B for unequal principal confinement frequencies. The potential is given by,

\begin{equation} V_{\mathrm{an}}(x,y) = \frac{\mu}{2} \left( \omega_x^2x^2+\omega_y^2y^2 \right), \label{eq:anisotropic_potential} \end{equation}

where $\mu$ is the effective mass and $\omega_x,\omega_y$ are the principal-axis frequencies ;with $\omega_x\neq\omega_y$ breaking rotational symmetry.

For a perpendicular field, the Hamiltonian is

\begin{equation} \hat{H} = \frac{\hat{p}_x^2+\hat{p}_y^2}{2\mu} -\frac{\omega_c}{2}\hat{L}_z + \frac{\mu}{2} \left( \Omega_x^2x^2+\Omega_y^2y^2 \right), \label{eq:anisotropic_hamiltonian} \end{equation}

with

\begin{equation} \Omega_x= \sqrt{\omega_x^2+\frac{\omega_c^2}{4}}, \qquad \Omega_y= \sqrt{\omega_y^2+\frac{\omega_c^2}{4}}, \label{eq:anisotropic_effective_frequencies} \end{equation}

where

\begin{equation} \omega_c=\frac{qB}{\mu}, \qquad \hat{L}_z=x\hat{p}_y-y\hat{p}_x . \end{equation}

For $\omega_x\neq\omega_y$, $\hat L_z$ is not conserved and the field couples the Cartesian directions; the states therefore lack definite magnetic quantum numbers. The quadratic Hamiltonian has exact normal-mode frequencies

\begin{equation} \omega_{\pm}^{2} = \frac{1}{2} \left[ \omega_x^2+\omega_y^2+\omega_c^2 \pm \sqrt{ \left(\omega_x^2+\omega_y^2+\omega_c^2\right)^2 -4\omega_x^2\omega_y^2 } \right], \label{eq:anisotropic_normal_modes} \end{equation}

and the corresponding energies are

\begin{equation} E_{n_+,n_-} = \hbar\omega_+ \left(n_++\frac{1}{2}\right) + \hbar\omega_- \left(n_-+\frac{1}{2}\right), \qquad n_\pm=0,1,2,\ldots . \label{eq:anisotropic_energy} \end{equation}

These energies provide an analytical benchmark. In the isotropic limit, $\omega_x=\omega_y=\omega_0$,

\begin{equation} \omega_{\pm} = \Omega\pm\frac{\omega_c}{2}, \end{equation}

and the conventional Fock--Darwin spectrum is recovered.

The anisotropic Hermite basis uses separate principal-axis scaling parameters:

\begin{equation} \alpha_x= \sqrt{\frac{\mu\Omega_x}{\hbar}}, \qquad \alpha_y= \sqrt{\frac{\mu\Omega_y}{\hbar}} . \label{eq:anisotropic_scaling} \end{equation}

The separate scales follow intrinsic anisotropy and magnetic localization. Using Ref.~\cite{DAHLIAH2026418388}, the InAs parameters are $\mu=0.0239m_e$, $\hbar\omega_x=6.0$~meV, and $\hbar\omega_y=5.6$~meV; the confinement lengths are
\begin{equation} \ell_x=\sqrt{\frac{\hbar}{\mu\omega_x}}, \qquad \ell_y=\sqrt{\frac{\hbar}{\mu\omega_y}}, \end{equation}

The lengths are $\ell_x\approx23.1$ and $\ell_y\approx23.9$~nm. Calculations extend to $10$~T to test basis convergence and spectral agreement with the normal-mode reference as magnetic coupling modifies the effective anisotropy.

\subsubsection{Computational Domain and Basis Convergence}
\label{sec:InAs_basis_convergence}

\begin{figure}[H]
    \centering
    \includegraphics[width=\textwidth]{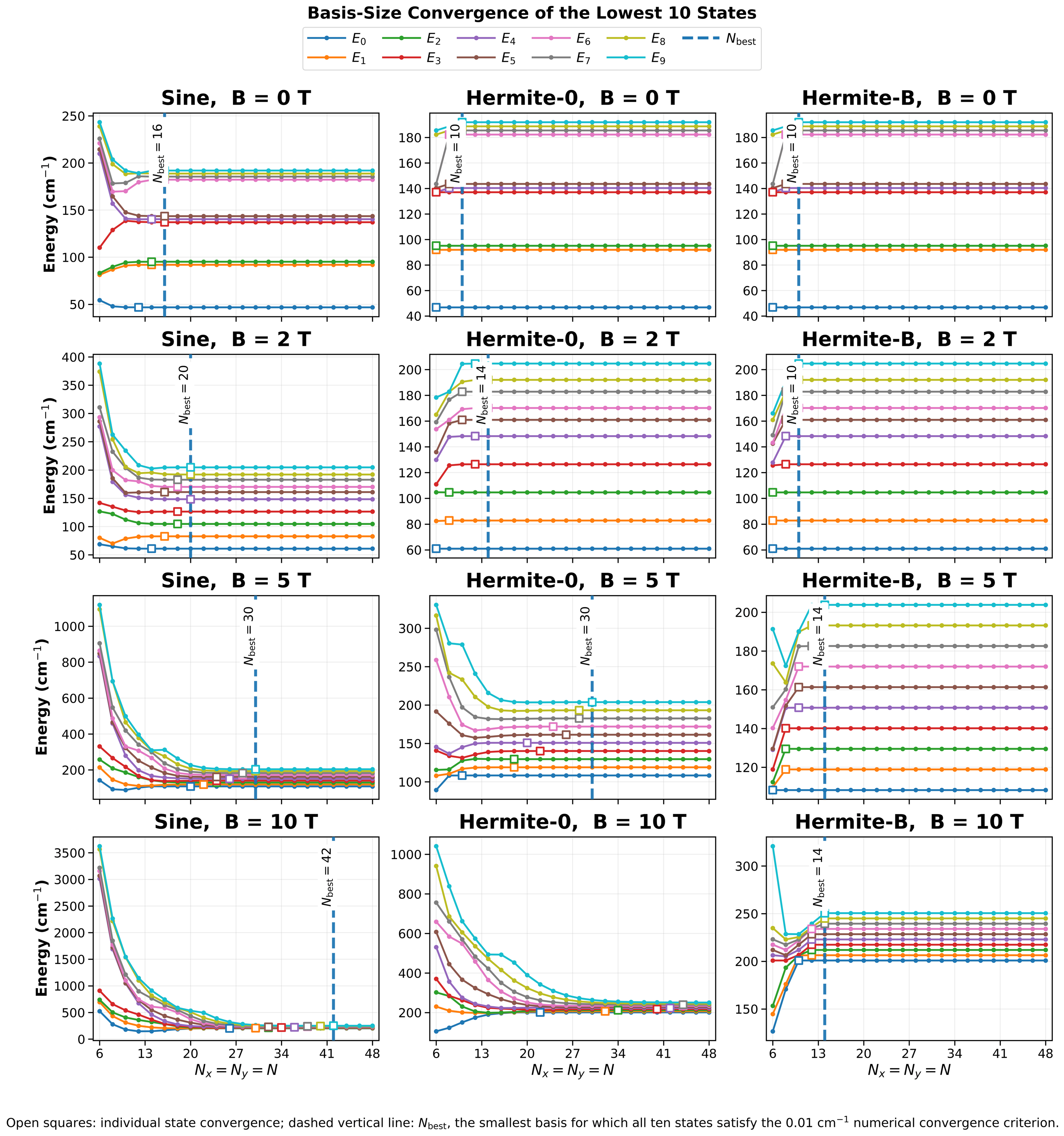}
    \caption{
Basis-size convergence of the lowest ten eigenstates for Sine, Hermite-0, and field-dependent Hermite-B DVRs for the anisotropic InAs quantum dot at $B=0$, $2$, $5$, and $10$~T. Open squares mark individual-state convergence at $0.01~\mathrm{cm}^{-1}$, while the dashed line indicates $N_{\rm conv}$ for simultaneous convergence of all ten states.
}
    \label{fig:basis_convergence_4x3}
\end{figure}

The fixed Sine-DVR domain spans approximately six characteristic confinement lengths per direction. Figure~\ref{fig:basis_convergence_4x3} reports convergence of the lowest ten states at $0.01$~cm$^{-1}$ for $B=0$, 2, 5, and 10~T. The simultaneous-convergence dimensions are $N_{\rm conv}=16$ (Sine) and 10 (both Hermite variants); open squares mark individual-state convergence and the dashed line the dimension converging all ten. Sine needs a larger basis before higher states stabilize. Hermite representations converge faster, although Hermite-0 and Sine become comparable at 5~T and Sine is more efficient at 10~T. Hermite-B remains compact over 0--10~T, with maximum $N_{\rm conv}=14$.

\subsubsection{Field-Dependent Spectrum and State Tracking}
\label{sec:state_tracking}

\begin{figure}[H]
    \centering
    \includegraphics[width=\textwidth]{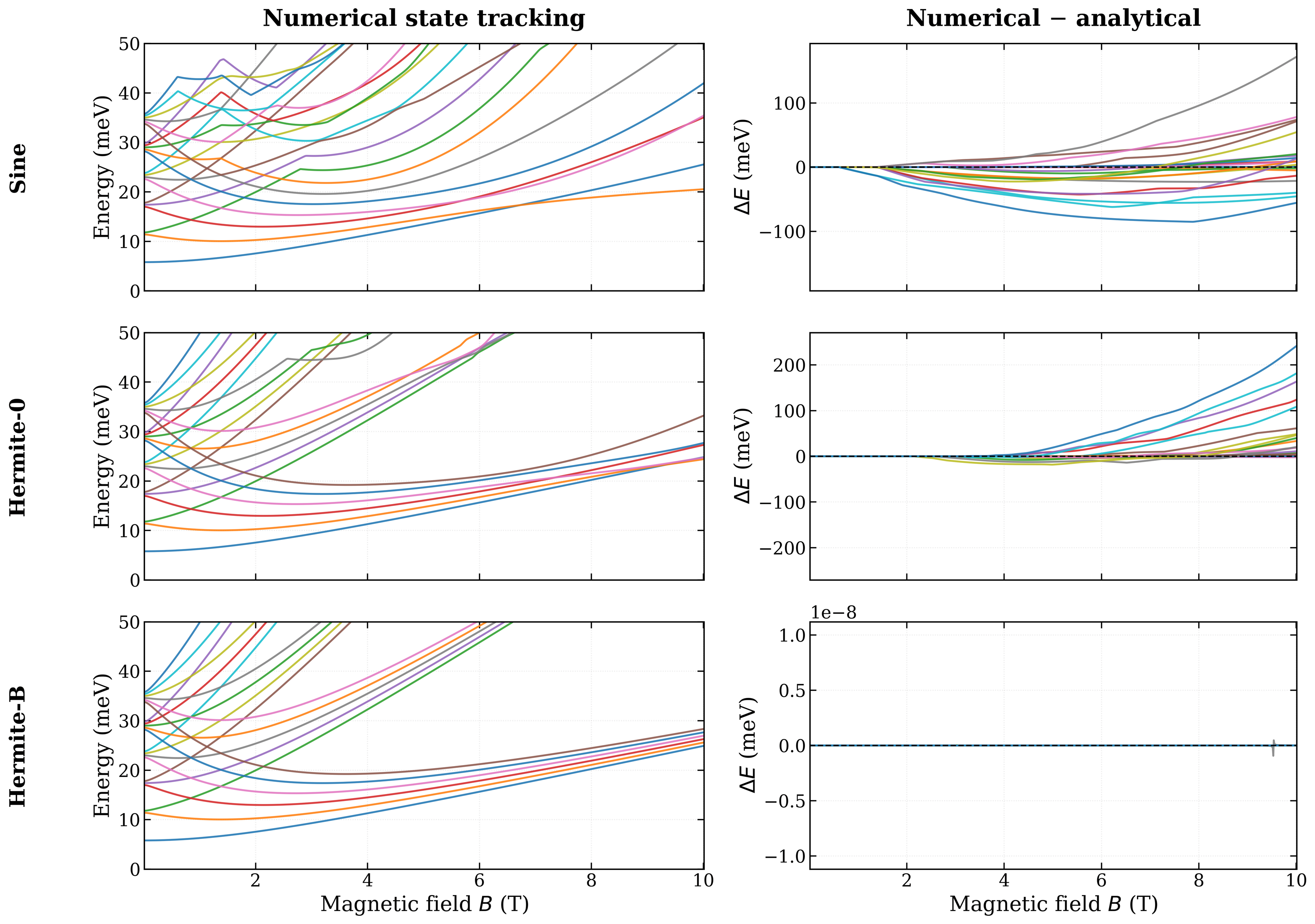}
    \caption{Field-dependent energy spectra of the first 21 states of the anisotropic InAs quantum dot. Rows correspond to Sine-DVR, Hermite-0, and Hermite-B; columns show state-tracked energies and deviations from analytical results.}
    \label{fig:field_spectrum_tracking}
\end{figure}

Figure~\ref{fig:field_spectrum_tracking} compares the first 21 states with the analytical spectrum using $N=16$ (the stated Hermite-B convergence dimension). Sine deviates visibly, especially for excited states and strong fields; Hermite-0 improves the spectrum but retains branch-specific discrepancies. Hermite-B follows the analytical branches closely, particularly at low energy, because its scale tracks the field-dependent spatial and energetic character of the states.

To preserve state identity when branches approach or exchange order, neighbouring-field eigenvectors were matched by overlap, with a small energy-weighting term and Hungarian assignments~\cite{Kuhn1955,Munkres1957} as discussed in Sec.~\ref{sec:NumericalImplementation}. This post-processing does not alter the independently diagonalized eigenvalues or eigenvectors. Discontinuities and larger errors in Sine and Hermite-0, visible even on the state energy scale, indicate basis limitations possibly affecting tracking accuracy. Hermite-B errors remain below $10^{-8}$~cm$^{-1}$.

\subsection{Isotropic Anharmonic Confinement: Woods--Saxon Potential}
\label{sec:WS_results}
\label{sec:woods_saxon}

Next, we test a non-parabolic finite confinement using the radially symmetric Woods--Saxon potential, a smooth finite-depth profile used for semiconductor nanostructures, quantum dots, and mesoscopic systems\cite{quinn2005composition,rowland1998photomodulated,Costa1999,xie2009study,Woods1954,lu2011effects,aghchegala2010effect,aghchegala2011effect,khaledi2014effect,dakhlaoui2023gaas,aytekin2013nonlinear}. Unlike the harmonic model, it has no analytical solution even in a perpendicular field. It is defined as

\begin{equation} V(r)=\frac{V_0} {1+\exp\left[(R_0-r)/\zeta\right]}, \label{eq:woods_saxon} \end{equation}

\begin{figure}[H]
    \centering
    \includegraphics[
        width=0.7\linewidth
    ]{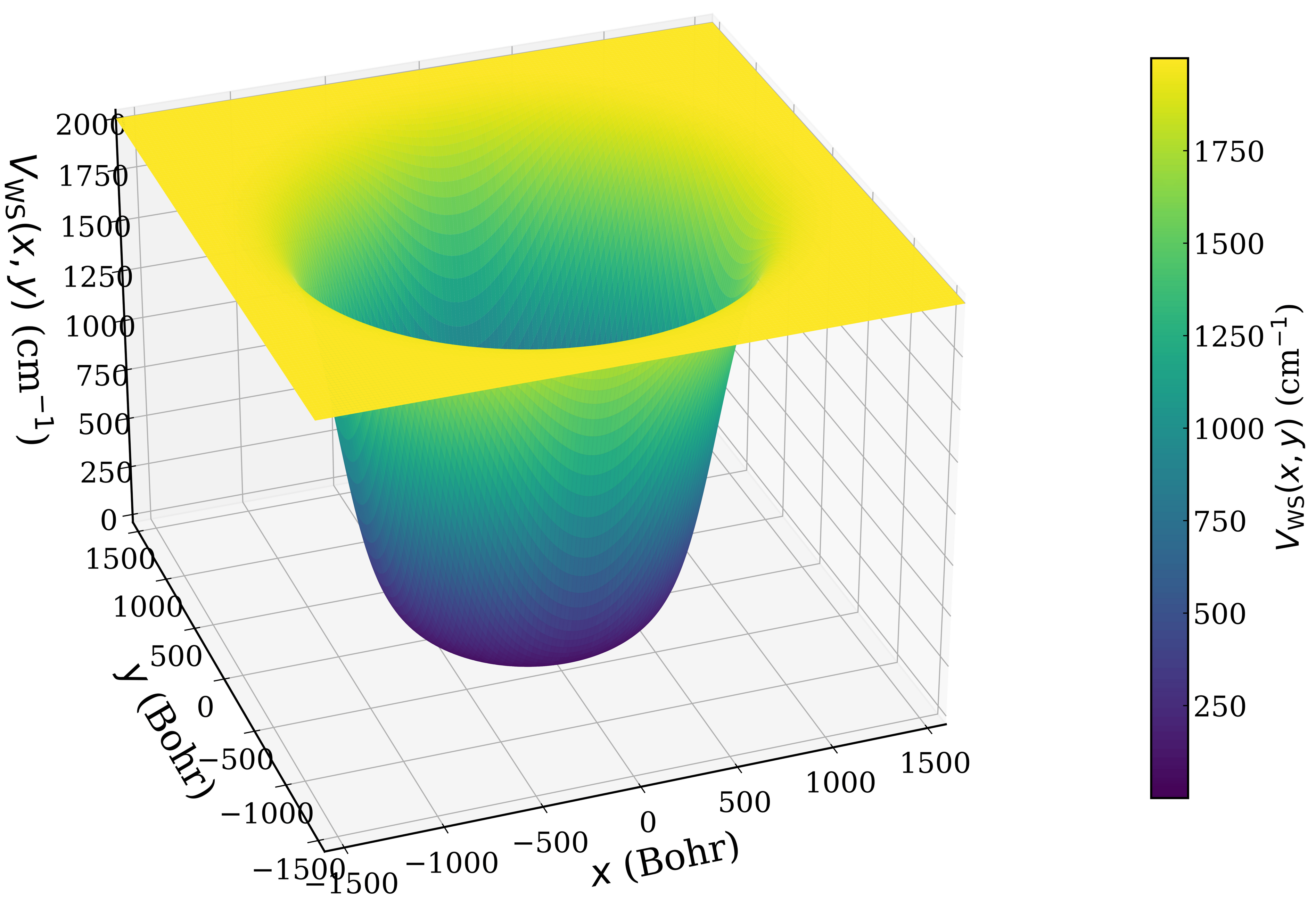}
    \caption{
        Three-dimensional representation of the Woods--Saxon confinement
        potential, $V(x,y)=V_0/\{1+\exp[(R_0-\sqrt{x^2+y^2})/\zeta]\}$.}
    \label{fig:woods_saxon_potential_3d}
\end{figure}

Here $V_0$ is the well depth, $R_0$ its characteristic radius, and $\zeta$ controls boundary diffuseness: $\zeta\to0$ gives a finite square well, while increasing $\zeta$ smooths the interface. The potential is shown in Fig.~\ref{fig:woods_saxon_potential_3d}. Its finite asymptote permits the wavefunction to extend beyond the nominal dot radius, requiring the basis to describe both the interior structure and the decaying tail. This therefore tests the Hermite-B representation beyond parabolic confinement.

Without an analytical reference, we compare basis convergence, spectral evolution, spatial spread, and computational cost. Since the Hermite-B scale is not predetermined for this potential, we first establish its selection protocol (Section~\ref{sec:WS_parameters}).

\subsubsection{Selection of Basis Scaling parameter for Woods--Saxon Confinement}
\label{sec:WS_parameters}
The radial limit $r^f_{\max}$ is defined where the potential reaches a fraction $f$ of its asymptotic value:
\begin{equation} V(r^f_{\max})=fV_0, \end{equation}
which gives
\begin{equation} r^f_{\max} = R_0+\zeta \ln\left(\frac{f}{1-f}\right). \label{eq:length} \end{equation}
For $V_0=0.00911$~a.u. ($2000$~cm$^{-1}$), $\zeta=94.486$~a.u. (5~nm), $R_0=944.86$~a.u. (50~nm), and $f=0.999$, $r^f_{\max}\approx1597.46$~a.u.; the Sine-DVR box length is
\begin{equation} L=2r^f_{\max}\approx 3194.92~\mathrm{a.u.} \end{equation}

Hermite DVR has no fixed box; its outermost quadrature point (the largest positive eigenvalue of the one-dimensional coordinate matrix) depends on $N$ and $\alpha$. We choose $\alpha$ by matching this extent to $r^f_{\max}$. At fixed $N$, varying $\alpha$ gives
\begin{equation} r_{\max}(\alpha;N) = \frac{C(N)}{\alpha}, \label{eq:rmax_alpha} \end{equation}
where $C(N)$ depends only on basis dimension (Table~\ref{tab:rvsa}).

\begin{table}[H]
    \centering
    \begin{tabular}{ccccc}
    \toprule
    S. No. & $N$ & Fitted relation for $r_{\max}(\alpha;N)$
    & $R^2$ & $C(N)$ \\
    \midrule
    1  & 8   & $2.9306\,\alpha^{-1}$  & 1 & 2.9306 \\
    2  & 16  & $4.6887\,\alpha^{-1}$  & 1 & 4.6887 \\
    3  & 20  & $5.3870\,\alpha^{-1}$  & 1 & 5.3870 \\
    4  & 24  & $6.0159\,\alpha^{-1}$  & 1 & 6.0159 \\
    5  & 28  & $6.5916\,\alpha^{-1}$  & 1 & 6.5916 \\
    6  & 32  & $7.1258\,\alpha^{-1}$  & 1 & 7.1258 \\
    7  & 40  & $8.0988\,\alpha^{-1}$  & 1 & 8.0988 \\
    8  & 48  & $8.9753\,\alpha^{-1}$  & 1 & 8.9753 \\
    9  & 56  & $9.7793\,\alpha^{-1}$  & 1 & 9.7793 \\
    10 & 64  & $10.526\,\alpha^{-1}$   & 1 & 10.526 \\
    11 & 80  & $11.888\,\alpha^{-1}$   & 1 & 11.888 \\
    12 & 96  & $13.116\,\alpha^{-1}$   & 1 & 13.116 \\
    13 & 128 & $15.932\,\alpha^{-1}$   & 1 & 15.932 \\
    \bottomrule
    \end{tabular}
    \caption{Dependence of the outermost Hermite-DVR grid point on the
    scaling parameter $\alpha$ for different numbers of basis functions.
    The fitted coefficient $C(N)$ characterizes the spatial extent of the
    corresponding Hermite-DVR grid.}
    \label{tab:rvsa}
\end{table}

A power-law fit to the tabulated $C(N)$ values gives

\begin{equation} C(N)=0.9395\,N^{0.5814}, \label{eq:C_N} \end{equation}

with $R^2=0.99899$. Combining Eqs.~\ref{eq:rmax_alpha} and \ref{eq:C_N} yields the outermost grid point

\begin{equation} r_{\max}(N,\alpha) = \frac{0.9395\,N^{0.5814}}{\alpha}. \label{eq:fr_max} \end{equation}

Matching this extent to Eq.~\eqref{eq:length} gives

\begin{equation} \alpha = \frac{0.9395\,N^{0.5814}}{r^f_{\max}}. \label{eq:alpha_opt} \end{equation}

For Hermite-0, the field-independent oscillator frequency (Eq.~\ref{eq:alpha}) gives

\begin{equation} \alpha^{WS} = \sqrt{\frac{m\omega_0^{WS}}{\hbar}}. \label{eq:alpha_omega0} \end{equation}

Here $\omega_0^{WS}$ is the characteristic Hermite-basis frequency, obtained from Eqs.~\ref{eq:alpha_opt} and \ref{eq:alpha_omega0}:

\begin{equation} \omega_0^{WS} = \frac{\hbar}{m} \left( \frac{0.9395\,N^{0.5814}}{r^f_{\max}} \right)^2. \label{eq:omega0} \end{equation}

Using Eq.~\ref{eq:length}, this becomes

\begin{equation} \omega_0^{WS} = \frac{\hbar}{m} \left[ \frac{0.9395\,N^{0.5814}} {R_0+\zeta\ln\left(\dfrac{f}{1-f}\right)} \right]^2. \label{eq:omega0_WS} \end{equation}

For Hermite-B, the field enters through $\omega_c$ and the effective frequency

\begin{equation} \Omega^{WS} = \sqrt{ \left(\omega^{WS}_0\right)^2 + \frac{\omega_c^2}{4} }. \label{effective_frequency_WS} \end{equation}

giving the field-dependent scaling parameter

\begin{equation} \alpha^{WS}_B = \sqrt{\frac{m\Omega^{WS}}{\hbar}}. \label{eq:alpha_omega_WS} \end{equation}

Thus, the potential sets $r^f_{\max}$, $N$ determines $\omega^{WS}_0$, and the magnetic term yields $\Omega^{WS}$ and $\alpha^{WS}_B$. The procedure also applies to other scalar potentials with asymptotic tails.

\subsubsection{Basis Optimization}
\begin{figure}[H]
    \centering
    \includegraphics[width=0.6\linewidth]
    {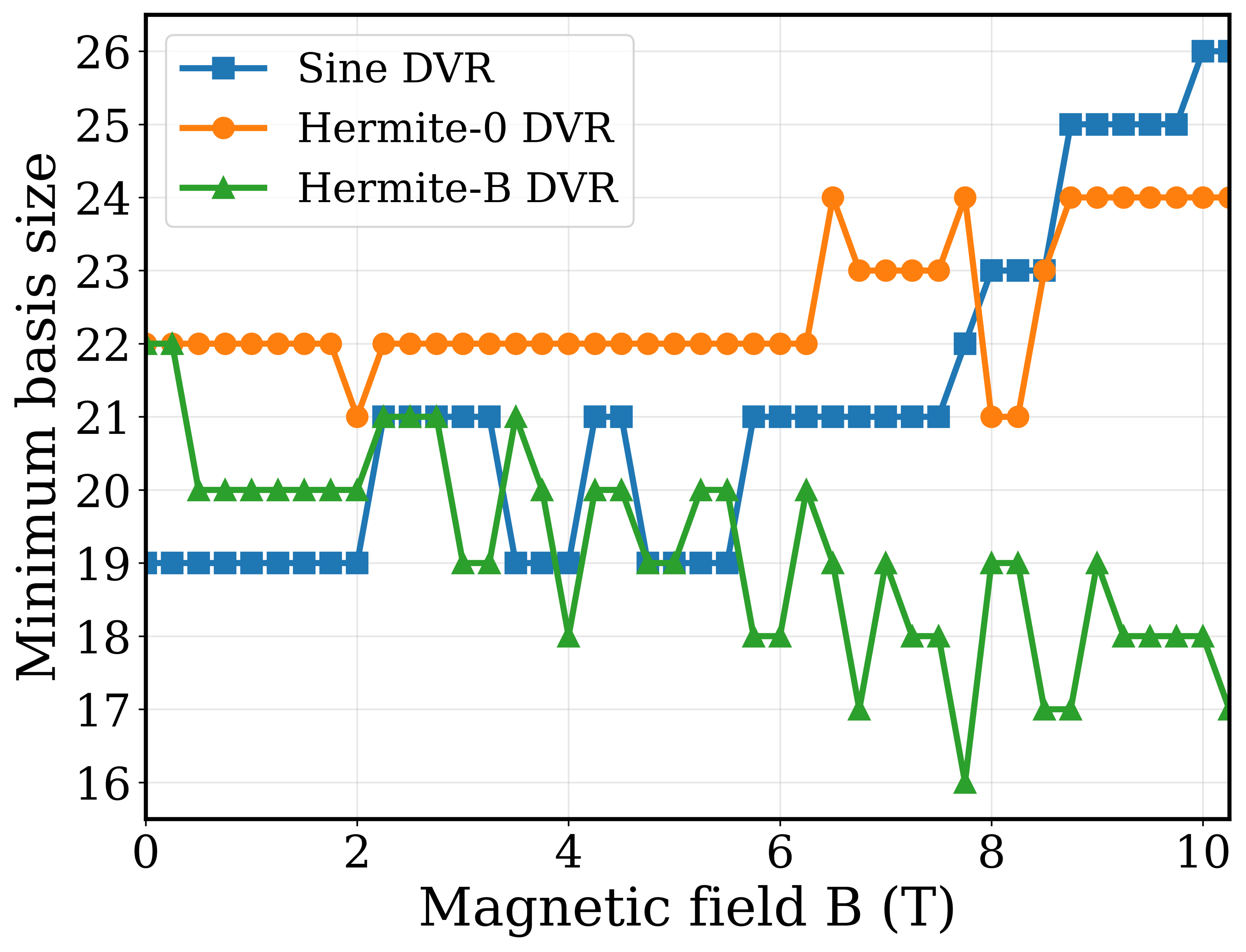}
    \caption{Minimum basis size required to achieve an energy convergence
    threshold of $10^{-2}$ cm$^{-1}$ for the sine DVR, Hermite-0 DVR,
    and Hermite-B DVR as a function of magnetic field.}
    \label{fig:basis_compare}
\end{figure}

Figure~\ref{fig:basis_compare} compares basis sizes at the $10^{-2}$~cm$^{-1}$ threshold, using the common extent $r^f_{\max}$.

At the lowest fields, Sine requires $N=19$ and both Hermite variants $N=22$, so Hermite-B has no weak-field advantage. Sine rises to $N=21$ at intermediate fields and then to 22, 23, 25, and 26 at the highest fields. Hermite-0 stays near 22, with isolated values of 24, 23, and 24. Hermite-B starts at 22, fluctuates mainly between 18 and 21, then reaches 17 and 16; above about 8~T it is predominantly in the range 17--19. These integer-threshold fluctuations reflect field-dependent state structure, for which small spectral changes can shift the minimum by one or more functions. Unlike Sine and Hermite-0, Hermite-B shows no systematic basis growth with field and uses about one-third fewer functions than Sine and 30\% fewer than Hermite-0 at the strongest fields. This also suggests that a zero-field-optimized Hermite-B basis can remain accurate at high fields.
 
\subsubsection{Field-Dependent Spectrum}
\label{sec:WS_energy_spectrum}

\begin{figure}[H]
    \centering
        \includegraphics[width=\linewidth]
        {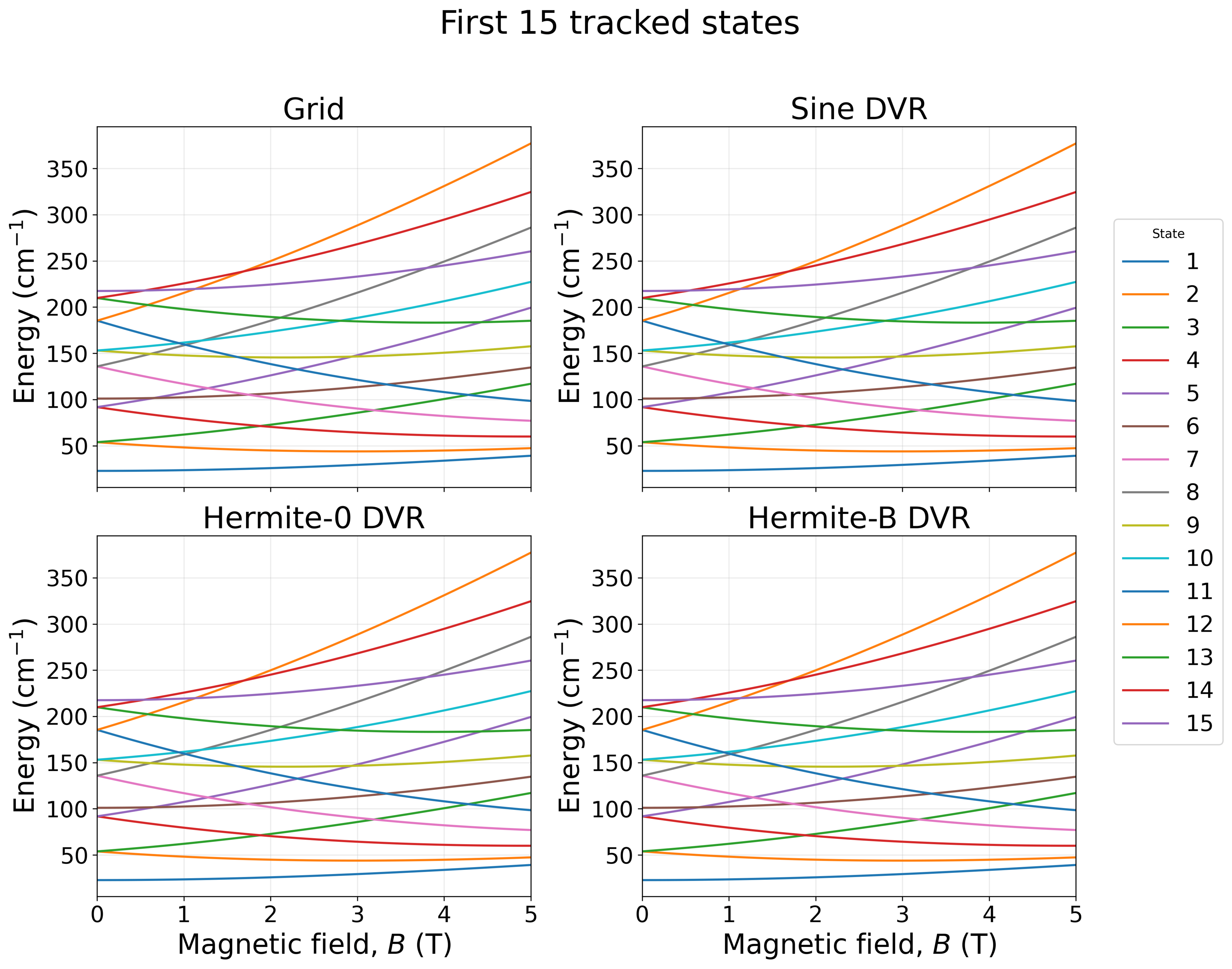}
    \caption{
    State--tracked Woods--Saxon energy spectrum for
    the grid and DVR methods studied. }
    \label{fig:WS_spectrum_and_errors}
\end{figure}

Figure~\ref{fig:WS_spectrum_and_errors} shows the tracked lowest ten states for the $512\times512$ Grid and all DVRs at $N=24$. Branch shifts and non-monotonic trends reflect competition between scalar and magnetic confinement. All methods reproduce the main splittings and curvature; deviations are $10^{-4}$--$10^{-3}$~cm$^{-1}$ among DVRs and $10^{-2}$--$10^{-1}$~cm$^{-1}$ for Grid, adequate for most applications.

\subsubsection{Spatial Representation and Grid Adaptation}
\label{sec:WS_spatial_representation}

We assess the ground-state spread $\Sigma_R$ and normalized spread $\Sigma_R/L_{\mathrm{eff}}$ as confinement increases.

\begin{figure}[H]
    \centering
    \includegraphics[width=0.9\linewidth]
    {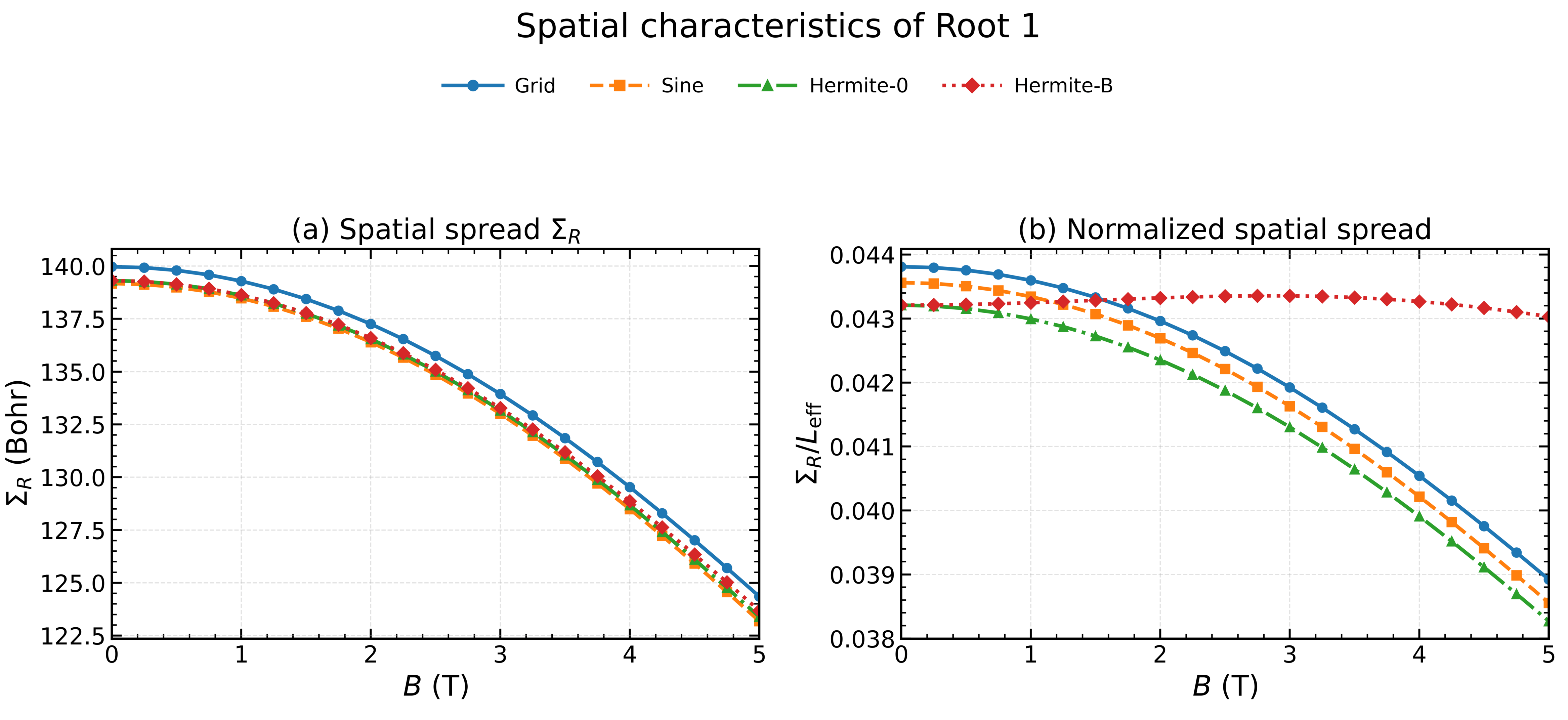}
    \caption{
    Spatial characteristics of the ground state as a function of
    magnetic field. (a) Spatial spread $\Sigma_R$, and (b) normalized
    spatial spread $\Sigma_R/L_{\mathrm{eff}}$
    }
    \label{fig:WS_spatial_analysis}
\end{figure}

The ground-state spread decreases from approximately 140 to 124~Bohr for all methods [Fig.~\ref{fig:WS_spatial_analysis}(a)]. The flatter Hermite-B normalized spread [Fig.~\ref{fig:WS_spatial_analysis}(b)] indicates more uniform sampling across the field range.


\subsubsection{Computational Efficiency}
\label{sec:WS_computational_efficiency}

\begin{figure}[H]
    \centering
    \includegraphics[width=0.95\linewidth]
    {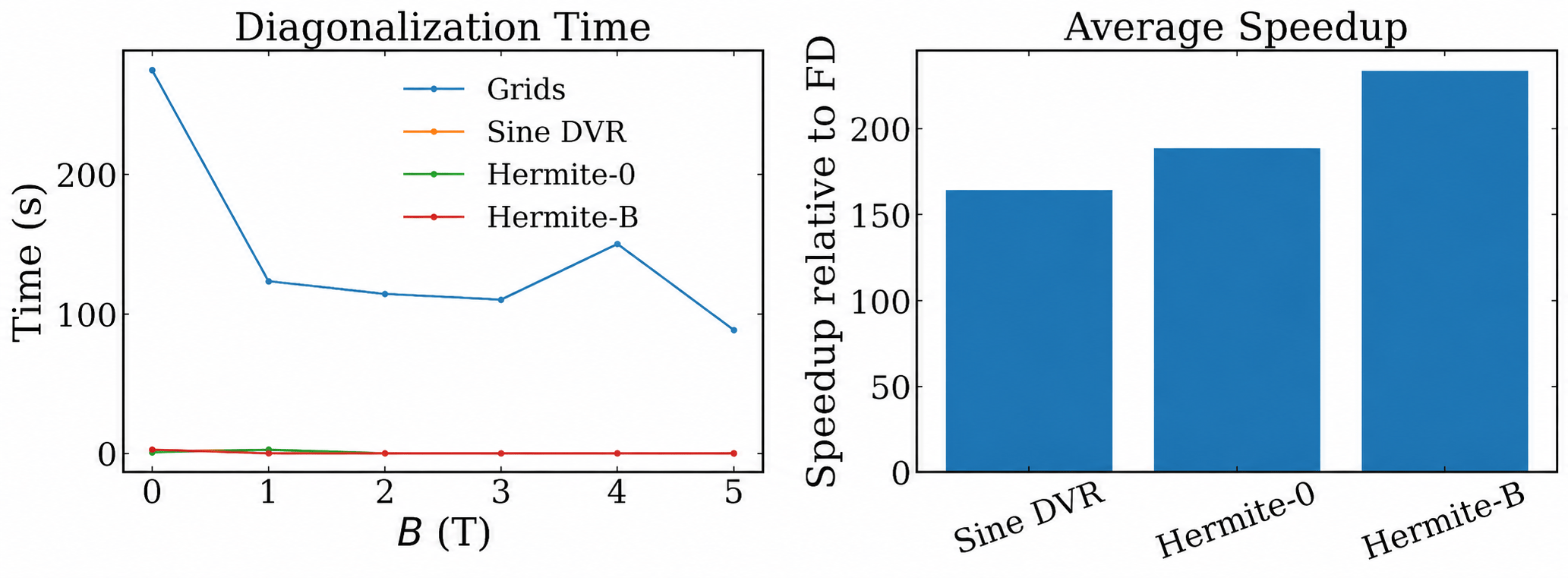}
    \caption{
    Computational cost of the grid (512$\times$512) and DVR calculations. The left panel shows the diagonalization time as a function of magnetic field, while the right panel shows the average speedup across $B=0-5 $ T of the DVR approaches relative to the finite-difference reference.
    }
    \label{fig:WS_computational_cost}
\end{figure}

Figure~\ref{fig:WS_computational_cost} compares diagonalization times for calculations used a 13th Gen Intel Core i7-13700 (24 logical CPUs, 128~GB RAM). The conventional grid Hamiltonian matrix was sparse, while DVR Hamiltonians were dense. 

DVR reaches maximum deviations of about $10^{-1}$~cm$^{-1}$ against analytical results with tens of basis functions, whereas Grid requires the $512\times512$ mesh for comparable accuracy.

Relative to Grid, Sine and Hermite-0 are about 100--200 times faster, and Hermite-B exceeds 200-fold speedup (same basis size $N=24$ for all DVRs). Thus, spectral accuracy can be restrained at lower cost with DVR methods for finite non-harmonic confinement.

\subsection{Arbitrarily Oriented Magnetic Fields}
\label{sec:Arbitrary_Field}

Arbitrarily oriented magnetic fields are relevant to magneto-optical spectroscopy and tilted-field transport in semiconductor nanostructures,\cite{peeters1990magneto,dippel1994charged,Hanson2007} as well as molecular ro-vibrational dynamics, where all field orientations relative to the freely rotating molecule must be considered.\cite{nikhily2025,yenugu2026} The present formulation accommodates such orientations without modifying the underlying DVR framework.

The magnetic field is expressed as
\begin{equation}
\mathbf{B} = B_x\hat{\mathbf i} + B_y\hat{\mathbf j} + B_z\hat{\mathbf k},
\label{vecB}
\end{equation}
where $\theta$ and $\phi$ denote its polar and azimuthal angles, respectively, with $\theta$ measured relative to the surface normal ($z$-axis). In the symmetric gauge,
\begin{equation}
\mathbf A = \frac12 \mathbf B \times \mathbf r.
\label{VecPot}
\end{equation}
For strict two-dimensional confinement in the $xy$ plane, the relevant in-plane projection of the vector potential is
\begin{equation}
\mathbf A_{\rm eff} =
\left(-\frac12B_z y,\frac12B_z x,0\right).
\label{effectVec}
\end{equation}
Using
\begin{equation}
B_x=B\sin\theta\cos\phi,\qquad
B_y=B\sin\theta\sin\phi,\qquad
B_z=B\cos\theta,
\end{equation}
the effective vector potential becomes
\begin{equation}
\mathbf A_{\rm eff} =
\left(-\frac12B\cos\theta\,y,
\frac12B\cos\theta\,x,0\right).
\label{effectVec_ang}
\end{equation}

The corresponding Hamiltonian,
\begin{equation}
\hat H =
\frac{1}{2\mu}(\hat{\mathbf p}-q\mathbf A_{\rm eff})^2
+\frac12\mu\omega_0^2r^2,
\end{equation}
can be expanded as
\begin{equation}
\hat H =
\frac{\hat p_x^2+\hat p_y^2}{2\mu}
-\frac{qB\cos\theta}{2\mu}\hat L_z
+\frac{q^2B^2\cos^2\theta}{8\mu}(x^2+y^2)
+\frac12\mu\omega_0^2(x^2+y^2).
\label{eq:tilted_H}
\end{equation}

Equation~(\ref{eq:tilted_H}) demonstrates that, for a particle of mass $\mu$ confined to the two-dimensional plane, the effect of an arbitrarily oriented magnetic field is equivalent to replacing the perpendicular magnetic field of Eq.~\ref{eq:alpha}by its normal component, $B\cos\theta$. Consequently, all operator matrices derived in the preceding sections remain unchanged, with only the effective magnetic field entering the Hamiltonian requiring modification.

For convenience in the discussion of the numerical results, we also introduce the complementary angle $\gamma=\pi/2 - \theta$, defined as the angle between the magnetic field and the plane of the quantum dot, so that

\[
B_z
=
B\cos\theta
=
B\sin\gamma.
\]

The numerical consequences of using a common basis across orientations are examined below, where we vary $\gamma$ while scaling Hermite-B with the full field magnitude $B$. In this strictly two-dimensional model, the parallel component produces no orbital confinement; the normal component alone controls the orbital response. This tests a common basis across orientations and motivates three-dimensional calculations in which field direction is an additional internal coordinate.

\subsubsection{Basis Convergence}

\begin{figure}[H]
\centering
\includegraphics[width=\linewidth]{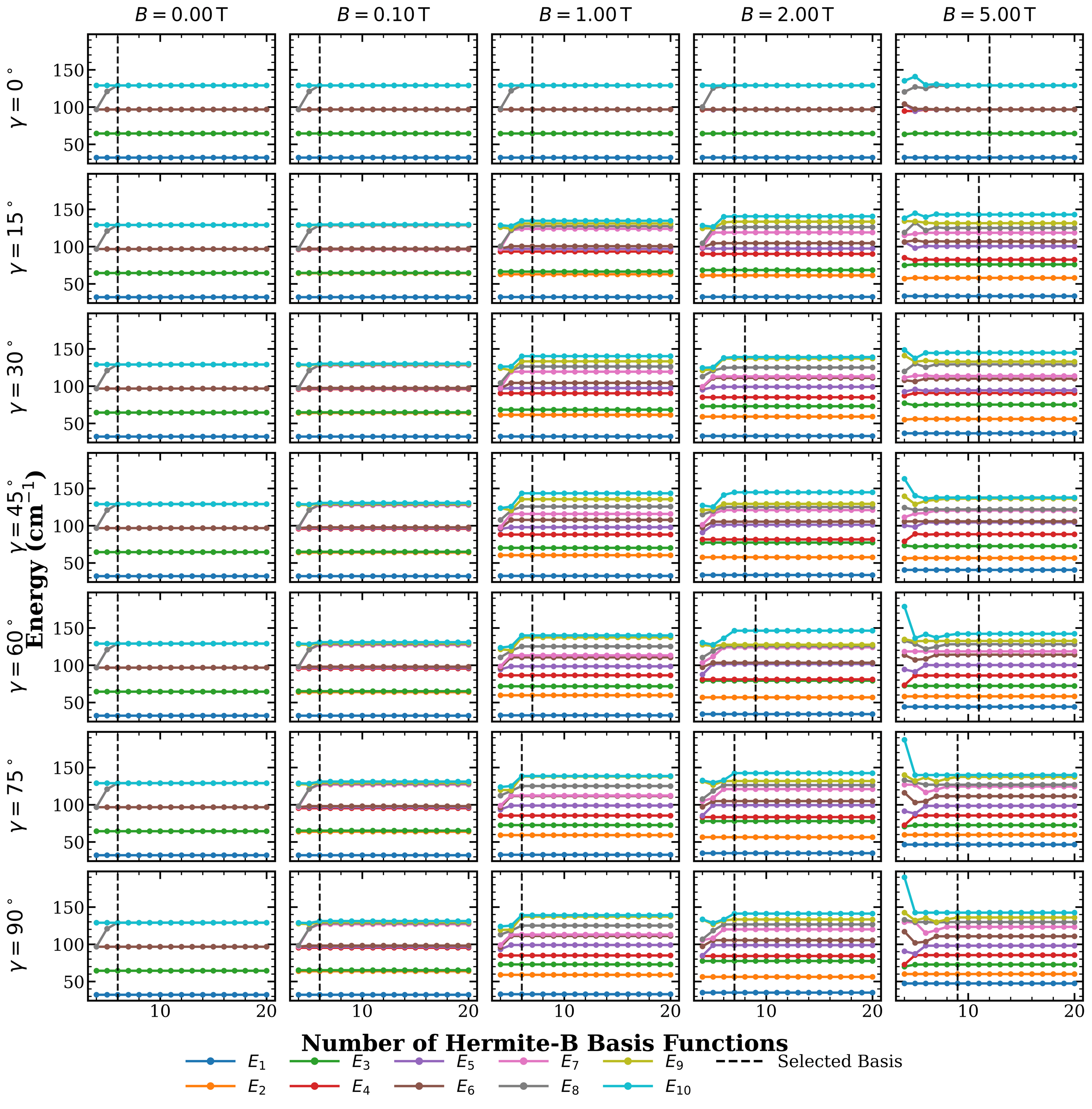}
\caption{Convergence of the lowest ten eigenvalues with respect to the number of Hermite-B basis functions for different magnetic-field strengths and magnetic-field orientations. The vertical dashed line indicates the selected basis dimension. $\gamma$ is the angle of the field $\vec{B}$ with the 2D plane.}
\label{fig:HermiteBasis}
\end{figure}

Figure~\ref{fig:HermiteBasis} shows convergence of the lowest ten eigenvalues for $B=0$--$5$~T and $\gamma=0^\circ$--$90^\circ$, using a common Hermite-B basis corresponding to the full $B$ at every angle. The dashed line marks the dimension satisfying the $10^{-2}$~cm$^{-1}$ criterion of Sec.~\ref{sec:NumericalImplementation}.

The lowest states converge rapidly, with selected dimensions varying by fewer than three functions over $\gamma=0^\circ$--$90^\circ$. Higher states are more sensitive, particularly at strong fields and large $\gamma$, and may converge non-monotonically. The criterion must therefore be checked for each $(B,\gamma)$ pair, with the final basis size selected as the largest number across all $(B,\gamma)$ and $n_{states}$ of interest.

\subsubsection{Evolution of the Energy Spectrum}

The tracked lowest-ten-state spectra (SI, Figs.~S145 and S146) vary smoothly with orientation. Excited-state field dependence weakens as $\gamma$ decreases, while the ground state changes only modestly; no numerical discontinuities or state-ordering changes occur.

\subsubsection{Wavefunction Analysis}

\begin{figure}[H]
    \centering
    \begin{subfigure}[b]{0.48\textwidth}
        \centering
        \includegraphics[width=\textwidth]{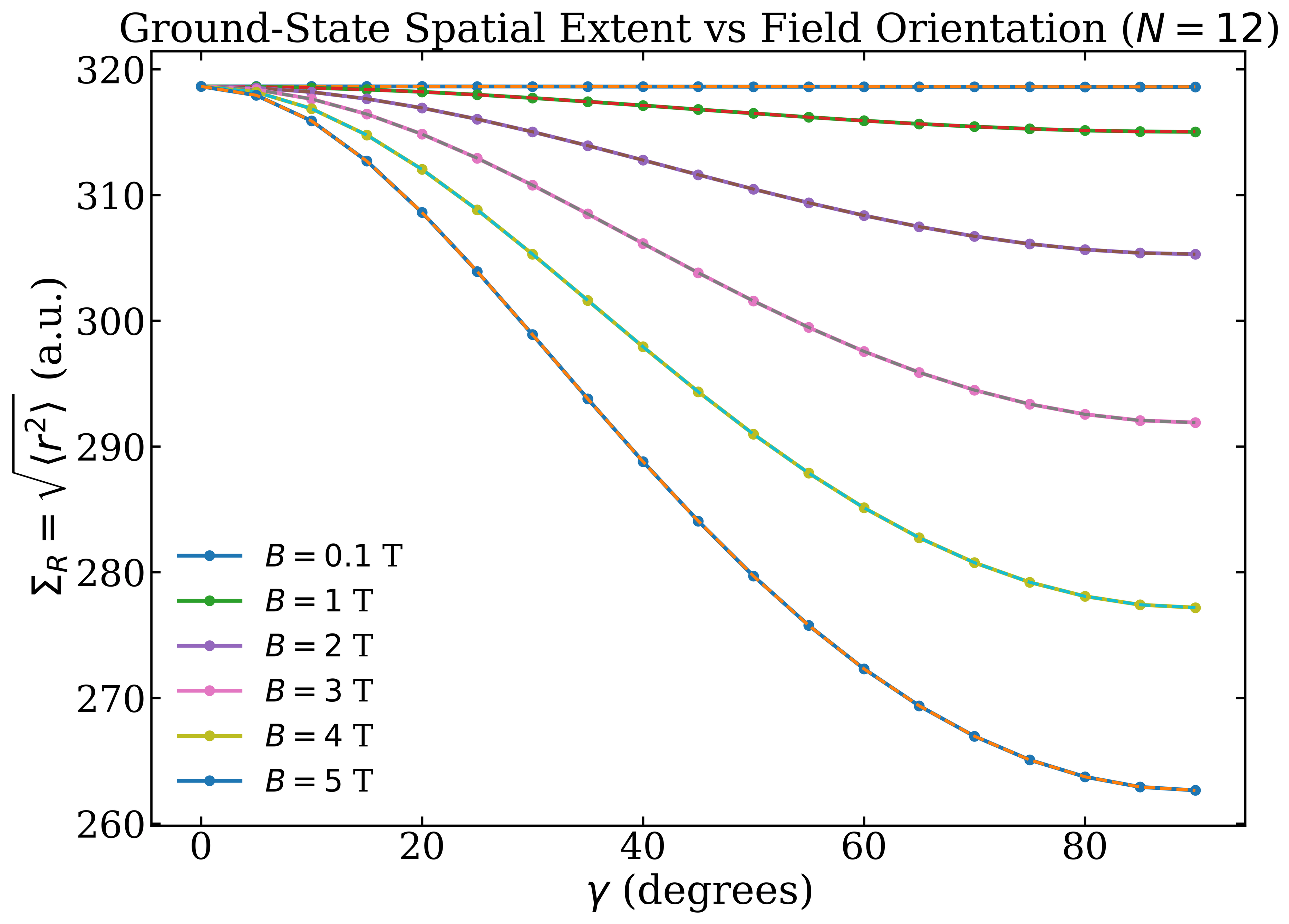}
        \caption{Ground-state spatial extent $\Sigma_R$ as a function of
        field orientation.}
        \label{fig:sigma_gamma}
    \end{subfigure}
    \hfill
    \begin{subfigure}[b]{0.48\textwidth}
        \centering
        \includegraphics[width=\textwidth]{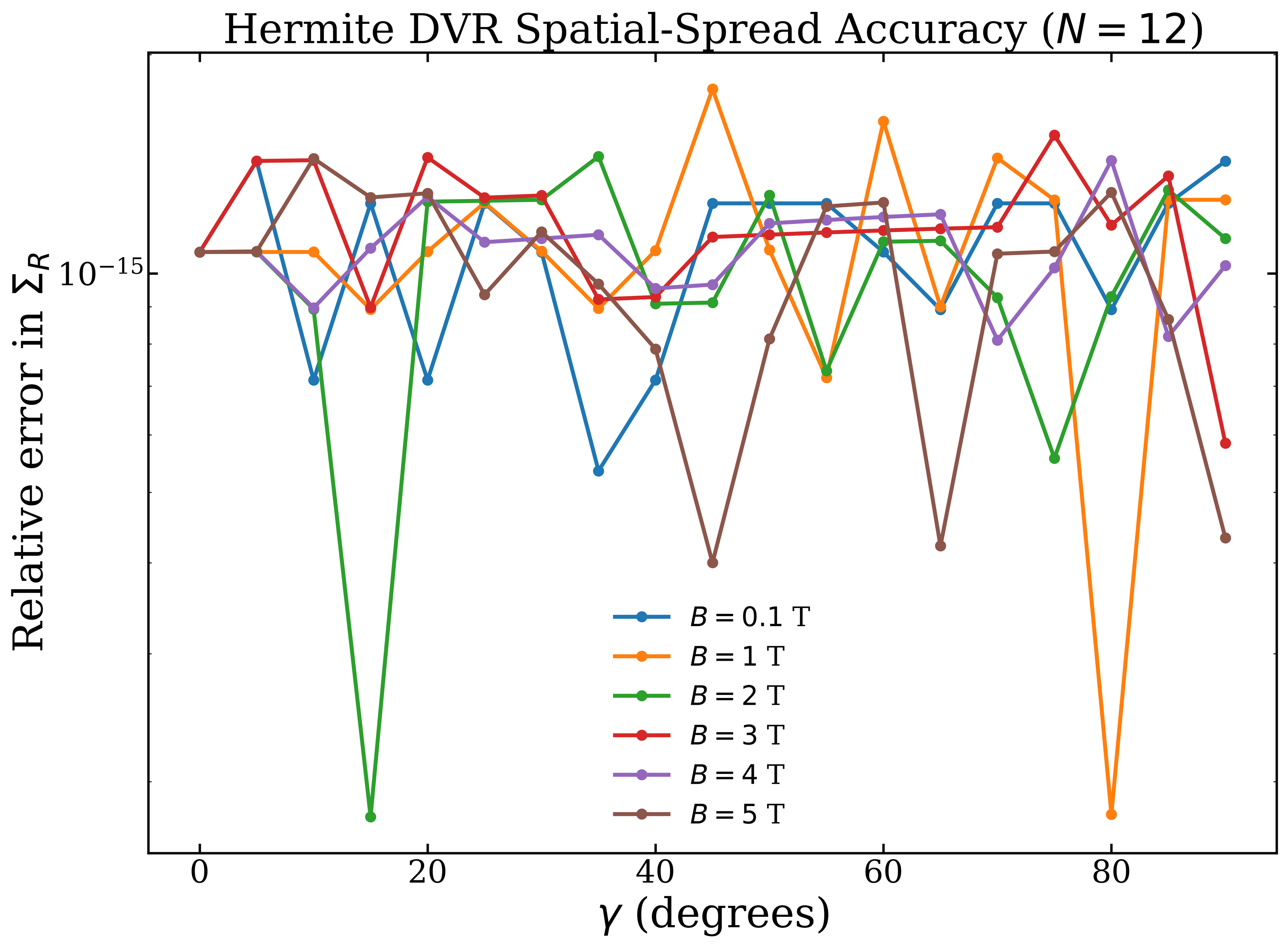}
        \caption{Relative error in the Hermite-DVR spatial extent with
        respect to the analytical Fock--Darwin result.}
        \label{fig:sigma_gamma_error}
    \end{subfigure}
    
    \caption{Orientation dependence of the ground-state spatial extent for
    different magnetic-field strengths. The field orientation is described
    by $\gamma$, where $\gamma=0$ and $\gamma=\pi/2$ correspond to magnetic
    fields parallel and perpendicular to the two-dimensional confinement
    plane, respectively.}
    \label{fig:comparison}
\end{figure}

Figure~\ref{fig:comparison}(a) shows $\Sigma_R=\sqrt{\langle r^2\rangle}$ versus $\gamma$. The curves approach a common value at $\gamma=0^\circ$, as expected without orbital coupling; increasing $\gamma$ contracts the ground state, especially at larger $B$. Hermite-B reproduces the analytical spread over the full angular range [Fig.~\ref{fig:comparison}(b)], with relative errors of order $10^{-15}$ even at strongest contraction. Energies and spatial structure are therefore described without angle-specific basis changes, supporting higher-dimensional applications with field orientation as an additional coordinate.

\section{Conclusions}
\label{sec:Conclusions}
In this paper, we explore, for the first time, DVR based computations of the eigen-spectrum of magnetically confined quantum dots. We find that the DVR-based methods provide a high accuracy ($\sim 10^{-1}$ cm$^{-1}$) at a much lower cost than finite-grid diagonalization. In particluar, we have developed a magnetic-field-dependent Hermite discrete variable representation (Hermite-B DVR) specifically for the accurate and efficient numerical treatment of magnetically confined quantum systems. The central feature of the formulation is the incorporation of the field-dependent confinement scale directly into the underlying Hermite basis. The numerical representation therefore follows the evolving physical length scale of the system rather than relying on a fixed representation over the entire field range.

The formulation was first established and benchmarked for a 2D isotropic harmonically confined GaAs quantum dot, against the Fock--Darwin solution and a 2D anisotropic harmonic InAs quantum dot, against its corresponding analytical solution. The Hermite-B DVR reproduces the analytical results with high accuracy while requiring substantially fewer basis functions than the corresponding field-independent representations, particularly in the stronger-field regime. 

The robustness of the approach was subsequently examined for a finite-depth Woods--Saxon confinement, where an analytical solution is not available. The DVR calculations reproduce the low-energy spectrum and spatial characteristics obtained from the finite-grid reference, while the Hermite-B representation retains its advantage in basis compactness as the magnetic confinement becomes stronger. The reduction in basis size translates into a corresponding reduction in computational effort, with the field-dependent Hermite representation providing the most favorable computational performance among the representations considered.

Finally, the field-dependent Hermite DVR was examined for continuously varying magnetic-field orientations. The same numerical framework accurately follows the evolution of both the energy spectrum and the spatial character of the eigenstates throughout the angular range, without requiring orientation-specific modifications of the basis. For the two-dimensional confinement considered here, the calculated results exhibit the expected dependence on the component of the magnetic field normal to the confinement plane. This demonstrates that the adaptive representation remains consistent not only under changes in magnetic-field magnitude, but also under continuous changes in field orientation.

The results establish that all the DVR methods improve on the grid based method in accuracy and cost while the principal advantage of the Hermite-B DVR is the incorporation of the evolving physical confinement scale into the numerical representation itself. By combining this adaptive character with the spectral accuracy of a basis representation and the local treatment of the potential inherent to DVR, the method provides a compact and computationally efficient description of multiple states across a broad range of magnetic-field strengths and confinement profiles.

\section*{Associated Content}
\subsection*{Supporting Information}

The Supporting Information (SI) is available free of charge. It contains additional calculations and datasets corresponding to the figures in the paper (ZIP File attached).

\begin{acknowledgement}
All authors gratefully acknowledge IISER Kolkata for infrastructure. P.B. acknowledges University Grants Commission (UGC) for his research fellowship. The authors acknowledge the use of ChatGPT (OpenAI) as an auxiliary tool for computational code development and debugging. We also acknowledge the National Supercomputing Mission (NSM) for providing computational resources at PARAM RUDRA, hosted at the S. N. Bose National Centre for Basic Sciences, implemented by C-DAC, and supported by the Ministry of Electronics and Information Technology (MeitY) and the Department of Science and Technology (DST), Government of India.
\end{acknowledgement}


\bibliography{JCTC_DVR}
\end{document}